\pdfoutput=1
\documentclass[
 reprint,
superscriptaddress,
preprintnumbers,
 amsmath,
 amssymb,
 aps,
 prapplied,
floatfix,
longbibliography
]{revtex4-2}
\usepackage{graphicx}

\usepackage{array}
\usepackage{braket}
\usepackage{mathtools}
\usepackage{physics}
\usepackage{amsmath,amsthm,amssymb,amsfonts}
\usepackage{booktabs}
\usepackage{hyperref}
\usepackage{pdfpages}
\usepackage{pgffor}

\makeatletter
\AtBeginDocument{\let\LS@rot\@undefined}
\makeatother

\newtheorem{theorem}{Theorem}
\newtheorem{lemma}{Lemma}

\newcommand{\avgT}[1]{\left\langle #1 \right\rangle_T}
\newcommand{\avginf}[1]{\left\langle #1 \right\rangle_\infty}

\makeatletter
\def\frontmatter@finalspace{\addvspace{10\p@}}
\makeatother

\begin{document}

\title{Classical Analog Emulation of Quantum Circuits\\
       via Time-Averaged Dynamic States}
\author{Matias~Miguel~Castillo~Valle}
\affiliation{Laboratory of Wave Engineering, EPFL, 1015 Lausanne, Switzerland}

\author{Tim~Tuuva}
\affiliation{Laboratory of Wave Engineering, EPFL, 1015 Lausanne, Switzerland}

\author{Benjamin~Apffel}
\affiliation{CNRS, ENS de Lyon, LPENSL, UMR5672, 69342 Lyon Cedex 07, France}

\author{Hervé~Lissek}
\affiliation{Laboratory of Wave Engineering, EPFL, 1015 Lausanne, Switzerland}

\author{Romain~Fleury}
\affiliation{Laboratory of Wave Engineering, EPFL, 1015 Lausanne, Switzerland}

\author{Mathieu~Padlewski}
\affiliation{Laboratory of Wave Engineering, EPFL, 1015 Lausanne, Switzerland}
\affiliation{Institut Langevin, ESPCI, CNRS, PSL Research University, 1 rue Jussieu, 75005, Paris, France}
\email{mathieu.padlewski@gmail.com}

\begin{abstract}
Between the digital and quantum paradigms lies the emerging field of quantum-inspired analog computing, in which classical waves emulate quantum circuits. We introduce \emph{wavebits} — classical wave analogs of qubits whose amplitudes are carried by physical oscillatory signals — and show that any nonseparable $N$-qubit state can be encoded in $2N$ narrowband signals that remain locally separable at every instant; the nonseparable correlations are recovered at readout by time-averaged demodulation over auxiliary carrier frequencies, the \emph{nonseparability channels}. We prove that any two-qubit gate \emph{unravels} into the time-averaged tensor product of two local time-varying operators, and that arbitrary circuits are emulated with a base-frequency count scaling linearly in the number of entangling two-qubit gates, independent of qubit number for a fixed gate count. The exponential cost of the $2^N$-dimensional state reappears at readout and in the averaging time of circuits with many two-qubit gates, not in circuit execution, as quantified by an analytic error bound that doubles as a hardware design rule. A mixed-signal prototype emulates Bell state generation, controlled-NOT gates, phase kickback, and Bloch-sphere rotations at fidelities above $0.99$, and numerical benchmarks against exact statevectors validate the scheme for up to six qubits. The architecture is directly implementable in acoustic, photonic, and mixed-signal platforms.
\end{abstract}
\maketitle

\section{Introduction}
\label{sec:intro}

Quantum hardware promises computational power beyond that of classical machines, but its fragility and cryogenic infrastructure keep scalable devices out of reach, while conventional digital computers are approaching fundamental limits of speed and energy efficiency. Quantum-inspired analog computing fills the space between these two paradigms: it uses the physics of waves to \emph{emulate} quantum circuits at the gate level on classical, room-temperature hardware, with benchmarking and teaching as natural by-products.  We use \emph{emulation} advisedly, in the sense articulated by La~Cour and Ott~\cite{lacour_signal-based_2015}: whereas a \emph{simulator} represents the quantum state numerically and computes matrix operations upon it in digital memory, an emulator physically instantiates an analogous dynamical structure (much as an inductive-capacitive circuit emulates a mass-spring system), so that the Hilbert-space amplitudes are carried by physical degrees of freedom and the gates are enacted by physical operations in continuous time.

The structure that any such emulator must reproduce is dictated by quantum information theory. Quantum entanglement is one of the most profound features of quantum mechanics, with applications spanning quantum cryptography~\cite{ gisin_quantum_2002,pirandola_advances_2020,bennett_quantum_2014,ekert_quantum_1991,scarani_security_2009} to quantum computation~\cite{feynman_simulating_1982,deutsch_quantum_1997, shor_algorithms_1994,nielsen_quantum_2010}.  The computational advantage of quantum computers fundamentally rests on \textit{nonseparability}: multipartite states that cannot be written as tensor products of individual subsystem states~\cite{karimi_classical_2015,korolkova_operational_2024}.  While nonlocality is a hallmark of quantum physics, it does not contribute directly to quantum speedup — nonseparability of the state space is the key resource~\cite{jozsa_role_2003,deutsch_harnessing_2020}.  The effort to reproduce this structure classically has a substantial history.

Spreeuw first recognized a formal analogy between classical optics and multi-qubit states~\cite{spreeuw_classical_1998,spreeuw_classical_2001}, and Kwiat \emph{et al.} implemented Grover's search with classical optical elements~\cite{kwiat_grovers_2000}; in both cases the physical resources scale exponentially with qubit number. In electronics, quadrature-modulation qubits~\cite{ferry_quantum_2001}, Hilbert-space analog computing~\cite{kish_quantum_2003}, and quantum-circuit processors~\cite{fujishima_16-qubit_2003} emulate the state explicitly, component by component. The analogy has even made its way to acoustics with the recent development of the phi-bit platform~\cite{hasan_sound_2019,hasan_experimental_2021,hasan_modeling_2022,ige_information_2024, deymier_practical_2024}.

Most directly relevant to the present work, La~Cour and Ott~\cite{lacour_signal-based_2015} demonstrated a signal-based classical emulation of a universal quantum computer in which the full $2^N$-dimensional state is encoded on a single signal as a nested sequence of amplitude-modulated \emph{tonals} (single-frequency sinusoidal components, one per computational basis state); gates are realized by analog electronics. Just as AM radio encodes distinct programs on unique carrier frequencies (each resolved by demodulation), each complex probability amplitude of the underlying $N$-qubit state is encoded here by a distinct tonal (a discrete frequency component). Using octave frequency spacing, the required bandwidth of the single signal grows exponentially with $N$: by the authors' own estimate, any practical electronic band, $0.1\,\mathrm{Hz}$--$100\,\mathrm{GHz}$, caps the approach at roughly forty qubits.  Moreover, because all qubits share one carrier, addressing them requires projection operations and pair-specific gate hardware. It is worth noting that subsequent work mitigates this bandwidth ceiling by offloading the frequency encoding \emph{spatially}, across parallel signals \cite{lacour_parallel_2018,mourya_emulation_2023}.

In this article, we propose an alternative encoding scheme that defers the exponential cost of the $2^N$-dimensional state to the final readout stage rather than paying it during circuit execution, substantially simplifying the analog mapping and overall circuit architecture. Our framework is built on a wave-native, multi-channel architecture equipped with a complete gate-level mapping. Here, we introduce a classical wave representation of a qubit, termed \emph{wavebit}, whose amplitudes are carried by physical oscillatory signals rather than static complex numbers. The time-average of a \emph{separable} composite wavebit state can reproduce the amplitudes of an \emph{arbitrary nonseparable} static quantum state, provided the amplitude signals are modulated by appropriately chosen nonseparability channels (NSCs): auxiliary angular frequencies, each acting as a dedicated carrier for one nonseparable correlation. To be clear, the construction is an encoding: at no instant is the physical system entangled. Its architectural advantage lies in how the encoding is distributed: unlike the single-signal scheme of Ref.~\cite{lacour_signal-based_2015}, whose bandwidth grows as $2^N$, the wavebit state is distributed over $2N$ separate narrowband channels (two per qubit) where the number of base frequencies (BFs) required during circuit execution scales \emph{linearly} in the number of entangling two-qubit gates $\delta$.  However, the exponential cost of the $2^N$-dimensional state does not disappear — as expected for any classical emulation of quantum circuits. It resurfaces at the \emph{readout stage}, where recovering the full output state requires $2^N$ time-averaged $N$-way signal products, and in the averaging time needed for circuits with many two-qubit gates.

This framework advances two concurrent agendas.  At the conceptual level, it provides a dynamical \emph{parameterization} of nonseparable states in terms of coherent multi-frequency classical dynamics, complementing established entanglement resource theories.
At the practical level, it furnishes a constructive recipe for the analog emulation of quantum circuits on classical wave hardware (Sec.~\ref{sec:discussion}).\\

\textit{Organization.} Section~\ref{sec:dynamic_states} introduces wavebits and demonstrates the Bell-state unraveling. Section~\ref{sec:generalization} derives the general $N$-qubit formula with NSCs.  Section~\ref{sec:gates} extends the framework to unitary operators and arbitrary circuits.  Section~\ref{sec:hardware} presents the mixed-signal analog hardware implementation and experimental validation. Section~\ref{sec:simulations} presents numerical simulations for up to $N=6$ qubits. Section~\ref{sec:discussion} relates the framework to quantum entanglement theory and to prior emulation schemes, details its limitations, and outlines the outlook.

\section{Wavebits and the Bell State}
\label{sec:dynamic_states}

\subsection{Wavebits and dynamic states}

In quantum information, a single qubit is described by a normalized state $\ket{\psi} = \alpha_0\ket{0} + \alpha_1\ket{1}$, where $\alpha_0,\alpha_1$ are complex-valued probability amplitudes.  We generalize this to a \emph{dynamic state},
\begin{equation}
    \ket{\psi(t)} = {\begin{pmatrix} \alpha_0(t)\\ \alpha_1(t)\end{pmatrix} = \alpha_0(t)\ket{0} + \alpha_1(t)\ket{1},}
    \label{eq:dynamic_state}
 \end{equation}
i.e., a column vector of two complex signals $\alpha_0(t)$ and $\alpha_1(t)$, taken to be finite superpositions of tones $e^{\mathrm{i}\omega t}$. Throughout this work, the explicit time dependence $(t)$ marks such classical signals and distinguishes them from the static quantum amplitudes (e.g., $\alpha_0(t)$ versus $\alpha_0$). Unlike conventional Hamiltonian evolution, where a unitary operator $e^{-\mathrm{i}Ht/\hbar}$ acts jointly on all amplitudes, here $\alpha_0(t)$ and $\alpha_1(t)$ evolve \emph{independently}, allowing the temporal degree of freedom to be used as a resource. We call $\ket{\psi(t)}$ a \emph{wavebit}: the classical wave representation of a dynamic state, in which the quantum amplitudes are carried by physical oscillatory signals rather than static complex numbers.  The corresponding \emph{static state} is recovered by temporal averaging:
\begin{equation}
    {\ket{\psi} = \begin{pmatrix}\alpha_0\\ \alpha_1\end{pmatrix}
    = \lim_{T\to\infty}\frac{1}{T}\int_0^T \begin{pmatrix} \alpha_0(t)\\ \alpha_1(t)\end{pmatrix}dt
    \coloneqq \avginf{\ket{\psi(t)}}.}
    \label{eq:time_avg}
\end{equation}
Any physical implementation averages over a finite window, $\avgT{f}\coloneqq\frac{1}{T}\int_0^T f(t)\,dt$, which approximates $\avginf{f}$ with an error bounded in Sec.~\ref{subsec:error}.

\textit{Note on normalization.}
The normalization condition $|\alpha_0|^2+|\alpha_1|^2=1$ applies to the time-averaged (static) state.  At each instant $t$, $|\alpha_0(t)|^2+|\alpha_1(t)|^2$ need \emph{not} equal unity.  The physical measurement statistics (Born-rule probabilities) are recovered from the time-averaged state (Sec.~\ref{subsec:limitations}).

\subsection{Separability and the role of time}
\label{subsec:separability}

For a two-qubit \emph{static} system, the most general pure state is
\begin{equation}
    \ket{\Psi} = \alpha_{00}\ket{00} + \alpha_{01}\ket{01}
               + \alpha_{10}\ket{10} + \alpha_{11}\ket{11},
    \label{eq:static_bipartite}
\end{equation}
with $\alpha_{ij}\in\mathbb{C}$.  The state is nonseparable if and only if
\begin{equation}
    \det\begin{bmatrix}\alpha_{00} & \alpha_{01}\\\alpha_{10} & \alpha_{11}\end{bmatrix} \neq 0.
    \label{eq:sep_condition}
\end{equation}
No product of \emph{static} single-qubit states, $(\alpha_0^{(1)}\ket{0}+\alpha_1^{(1)}\ket{1})\otimes(\alpha_0^{(2)}\ket{0}+\alpha_1^{(2)}\ket{1})$, yields the maximally nonseparable Bell state $\ket{\Phi^+} = \frac{1}{\sqrt{2}}(\ket{00}+\ket{11})$.

With \emph{dynamic} amplitudes, however, the situation changes.  Forming the tensor product of two wavebits $\ket{\psi^{(n)}(t)}=\alpha_0^{(n)}(t)\ket{0}+\alpha_1^{(n)}(t)\ket{1}$, $n=1,2$,
\begin{multline}
    \ket{\Psi(t)} = \ket{\psi^{(1)}(t)}\otimes\ket{\psi^{(2)}(t)}\\
    = \underbrace{\alpha_0^{(1)}(t)\alpha_0^{(2)}(t)}_{\alpha_{00}(t)}\ket{00}
    + \underbrace{\alpha_0^{(1)}(t)\alpha_1^{(2)}(t)}_{\alpha_{01}(t)}\ket{01}\\
    + \underbrace{\alpha_1^{(1)}(t)\alpha_0^{(2)}(t)}_{\alpha_{10}(t)}\ket{10}
    + \underbrace{\alpha_1^{(1)}(t)\alpha_1^{(2)}(t)}_{\alpha_{11}(t)}\ket{11},
    \label{eq:dyn_product}
\end{multline}
the composite state $\ket{\Psi(t)}$ is separable at every instant $t$ while its time average can be nonseparable.  We call this \emph{unraveling} a nonseparable state.  Note that the average is taken over the amplitude signals — a coherent demodulation step rather than a physical operation on quantum states; the precise relation of this encoding to quantum entanglement is examined in Sec.~\ref{sec:discussion}.\\

\subsection{Unraveling the Bell state}

\begin{figure}[t]
\centering
\includegraphics[width=\columnwidth]{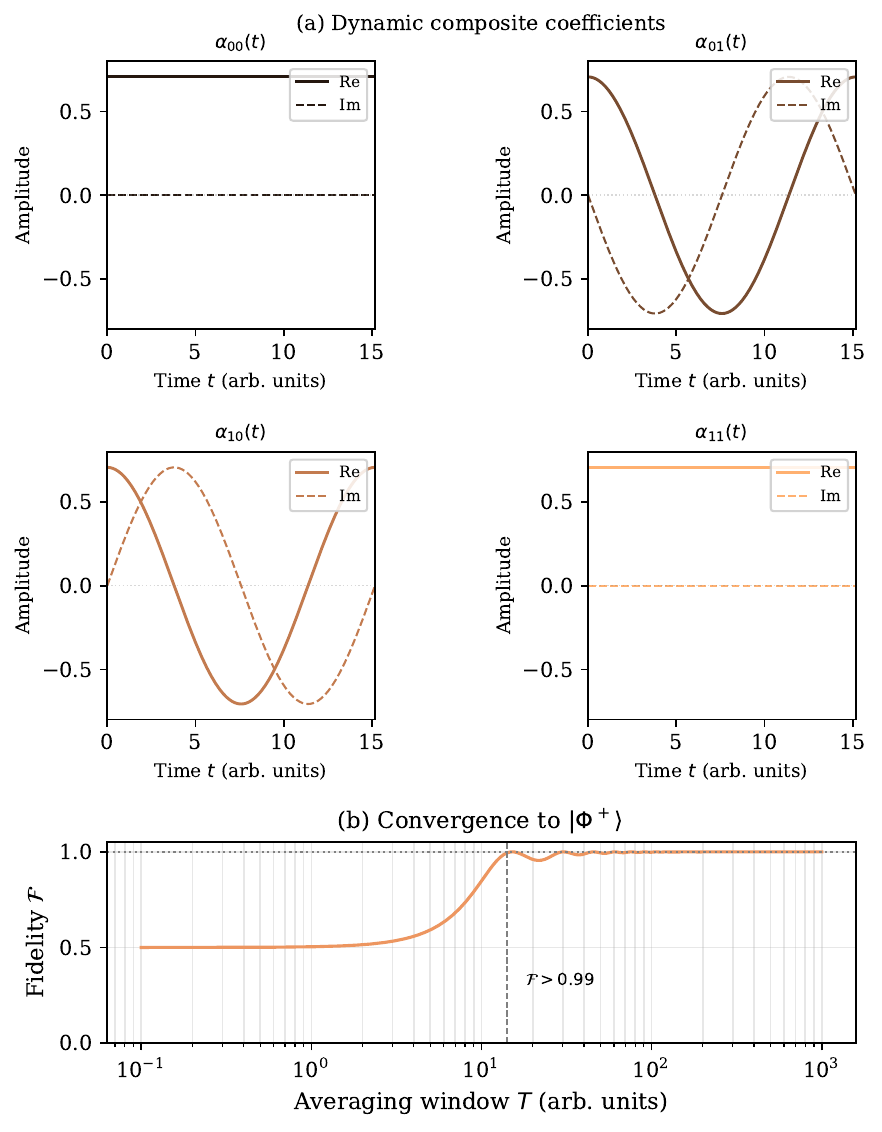}
\caption{\textbf{Bell-state unraveling via nonseparability channels.} (a)~Real (solid) and imaginary (dashed) parts of all four dynamic composite coefficients $\alpha_{ij}(t)$ with $\omega_B=\sqrt{2}\,\omega_A$: the diagonal terms $\alpha_{00}(t)$ and $\alpha_{11}(t)$ are constant and carry the entangled weight $1/\sqrt{2}$, while the off-diagonal terms $\alpha_{01}(t)$ and $\alpha_{10}(t)$ oscillate at $\pm(\omega_B-\omega_A)\neq 0$ and vanish upon time-averaging. (b)~Fidelity $\mathcal{F}$ of the time-averaged state with $\ket{\Phi^+}$ vs averaging window $T$ (log scale), for $\omega_A=1$ (arb.\ units). Convergence above $99\%$ is first reached at the dashed vertical marker.}
\label{fig:bell}
\end{figure}

To illustrate, consider the assignment
\begin{equation}
    {\alpha_0^{(1)}(t)} = 2^{-1/4}e^{+\mathrm{i}\omega_A t},\quad
    {\alpha_1^{(1)}(t)} = 2^{-1/4}e^{+\mathrm{i}\omega_B t},
    \label{eq:Bell_q1}
\end{equation}
\begin{equation}
    {\alpha_0^{(2)}(t)} = 2^{-1/4}e^{-\mathrm{i}\omega_A t},\quad
    {\alpha_1^{(2)}(t)} = 2^{-1/4}e^{-\mathrm{i}\omega_B t},
    \label{eq:Bell_q2}
\end{equation}
where $\omega_A \neq \omega_B$ are two distinct carrier frequencies. The composite coefficient matrix becomes
\begin{equation}
    \begin{bmatrix}\alpha_{00}(t) & \alpha_{01}(t)\\\alpha_{10}(t) & \alpha_{11}(t)\end{bmatrix}
    = \frac{1}{\sqrt{2}}\begin{bmatrix}1 & e^{\mathrm{i}(\omega_A-\omega_B)t}\\
    e^{\mathrm{i}(\omega_B-\omega_A)t} & 1\end{bmatrix}.
    \label{eq:Bell_dyn_matrix}
\end{equation}
Time-averaging using $\avginf{e^{\mathrm{i}\omega t}}=0$ for $\omega\neq 0$ (for a finite window, $|\avgT{e^{\mathrm{i}\omega t}}|\le2/(|\omega|T)$, so that $T\gg 2\pi/|\omega_A-\omega_B|$ suffices) yields
\begin{equation}
    {\begin{bmatrix}\avginf{\alpha_{00}(t)} & \avginf{\alpha_{01}(t)}\\
    \avginf{\alpha_{10}(t)} & \avginf{\alpha_{11}(t)}\end{bmatrix}}
    = \frac{1}{\sqrt{2}}\begin{bmatrix}1 & 0\\0 & 1\end{bmatrix}.
    \label{eq:Bell_avg}
\end{equation}
Hence $\avginf{\ket{\Psi(t)}}$ $= \frac{1}{\sqrt{2}}(\ket{00}+\ket{11}) = \ket{\Phi^+}$. The two frequencies $\omega_A$ and $\omega_B$ compose the NSC for the Bell state.  Their spectral separation ensures cross-terms vanish upon averaging, leaving only the desired nonseparable correlations. Fig.~\ref{fig:bell} shows the time traces of all four dynamic composite coefficients [panel~(a)] and the numerical convergence of the fidelity $\mathcal{F}=|\braket{\Psi_{\mathrm{avg}}}{\Phi^+}|^2$ with increasing $T$ [panel~(b)].

\section{\textit{N}-Qubit Unraveling}
\label{sec:generalization}

Let us now consider an arbitrary $N$-qubit state.  The corresponding dynamic composite state is
\begin{equation}
    \ket{\Psi(t)} = \bigotimes_{n=1}^N \ket{\psi^{(n)}(t)}
    = \sum_{\mathbf{b}\in\{0,1\}^N} \prod_{n=1}^N \alpha^{(n)}_{b_n}(t)\,\ket{\mathbf{b}},
    \label{eq:N_qubit_dyn}
\end{equation}
where $\ket{\psi^{(n)}(t)} = \alpha_0^{(n)}(t)\ket{0}+\alpha_1^{(n)}(t)\ket{1}$. The target static state is $\ket{\Psi}=\sum_{\mathbf{b}}\alpha_\mathbf{b}\ket{\mathbf{b}}$, with the time-averaging constraint
\begin{equation}
    {\avginf{\alpha_\mathbf{b}(t)} = \alpha_\mathbf{b},
    \quad \alpha_\mathbf{b}(t)\coloneqq\prod_{n=1}^N \alpha_{b_n}^{(n)}(t),}
    \quad\forall\,\mathbf{b}\in\{0,1\}^N.
    \label{eq:constraint}
\end{equation}

It becomes evident here that the number of distinct nonseparable correlations grows exponentially with $N$.  For example, the 3-qubit state has 8 nonzero coefficients $\{\alpha_{000},\alpha_{100},\alpha_{010},\alpha_{001},\alpha_{110},\alpha_{101},\alpha_{011},\alpha_{111}\}$, each corresponding to a distinct nonseparable correlation.  To satisfy Eq.~\eqref{eq:constraint}, each correlation must be carried by a distinct NSC. For $N$ qubits, each of the $2N$ amplitude functions ($\alpha_0^{(n)}$ and $\alpha_1^{(n)}$ for $n=1,\ldots,N$) requires $2^N$ NSCs, giving a total of $N\cdot 2\cdot 2^N$ frequencies per target state.  This grows \emph{exponentially} with $N$, because the $N$-qubit construction attempts to span the full $2^N$-dimensional Hilbert space simultaneously.  This section thus establishes the \emph{naive baseline}: the exponential cost one incurs when directly unraveling the output state.

In the following section, we show how this exponential frequency overhead can be avoided using analog circuits based on \emph{active gates}: rather than synthesizing the signals of Eq.~\eqref{eq:constraint} directly (a direct construction, with its exponential cost, is given in Appendix~\ref{app:gate_unravel}), nonseparable states are built up by applying gates to an initial product state, exactly as in a quantum circuit.

\section{Circuit Unraveling using Active Gates}
\label{sec:gates}

\subsection{Thinking in terms of circuits}\label{subsec:circuits}

The previous section motivates a new approach based around NSCs. Unraveled states could be manipulated without explicit computation of the $2^N$ components of the Hilbert space in which the full state resides. As we will show, the \emph{signal-generation} cost of these manipulations scales polynomially, in fact linearly, with the number of entangling two-qubit gates. Two costs remain, and it is important to state them plainly. First, the output cannot be read directly: to recover the full output state, one must apply the tensor product and time-average in an additional \textit{readout stage}, whose hardware cost is exponential in $N$ (Sec.~\ref{sec:discussion}), although this stage is circuit-independent, and thus the exponential cost appears only once. Second, the averaging time required for a target accuracy grows with the number of layers through the density of near-resonant combination frequencies (Secs.~\ref{subsec:error} and~\ref{subsec:sim_circuits}). Neither cost is incidental: a classical emulator that avoids both would efficiently simulate universal quantum circuits, which is not expected to be possible. The framework's contribution is to isolate \emph{where} the exponential cost lives — at readout and in convergence, not in signal generation.

\textit{Initial state.} Circuits act on a separable input, which in the quantum-circuit convention is $\ket{0\cdots0}$. It is encoded by constant signals, $\alpha_0^{(n)}(t)=1$ and $\alpha_1^{(n)}(t)=0$ for every wavebit $n$; any product state $\bigotimes_n(\alpha_0^{(n)}\ket{0}+\alpha_1^{(n)}\ket{1})$ is likewise encoded by constant signals $\alpha_{b}^{(n)}(t)=\alpha_{b}^{(n)}$, which is the choice made in our hardware and simulations. The encoding is not unique, since any zero-mean component may be added to a signal: for instance, $\alpha_1^{(n)}(t)=e^{\mathrm{i}\Omega_0 t}$ for all $n$ also encodes $\ket{0\cdots0}$, provided that $\Omega_0$ is linearly independent over $\mathbb{Q}$ of the base frequencies introduced below, so that it cannot recombine into a spurious DC term.

\subsection{Single-qubit gates}
\label{subsec:1qubit_gate}

A single-qubit gate $\hat V$ acting on qubit $n$ is applied as a \emph{static} $2\times2$ matrix to the two channels of wavebit $n$,
\begin{equation}
    \begin{pmatrix} \alpha_0^{(n)}(t)\\ \alpha_1^{(n)}(t)\end{pmatrix}\;\longmapsto\;\hat V\begin{pmatrix} \alpha_0^{(n)}(t)\\ \alpha_1^{(n)}(t)\end{pmatrix},
    \label{eq:1qubit_gate}
\end{equation}
leaving all other wavebits untouched. Because $\hat V$ is constant, it commutes with time averaging, $\avginf{\hat V\ket{{\psi^{(n)}}(t)}}=\hat V\avginf{\ket{{\psi^{(n)}}(t)}}$; and because it acts on a single tensor factor, it maps the composite signal state $\bigotimes_m\ket{{\psi^{(m)}}(t)}$ to $(\hat I\otimes\cdots\otimes\hat V\otimes\cdots\otimes\hat I)\bigotimes_m\ket{{\psi^{(m)}}(t)}$ at every instant, so that the encoded amplitudes transform exactly as the quantum state does. Single-qubit gates therefore require neither a base frequency nor any oscillating signal; in hardware, they are implemented by static gains (Sec.~\ref{sec:hardware}). The same holds for any $2\times2$ matrix, unitary or not.

\subsection{Unraveling a 2-qubit gate}
\label{subsec:2qubit_gate}
\begin{figure}[h]
\centering
\includegraphics{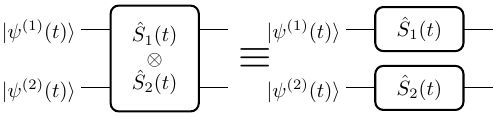}
\caption{\textbf{Separable structure of the gate unraveling.} The time-varying operator $\hat U(t) = \hat{S}_1(t)\otimes\hat{S}_2(t)$ (left) is equivalent to independent local operators $\hat{S}_1(t)$ and $\hat{S}_2(t)$ acting on each subsystem (right). The time-averaged tensor product of these locally separable operations recovers the target two-qubit gate $\hat{U}=\avginf{\hat{S}_1(t)\otimes\hat{S}_2(t)}$.}
\label{fig:gate_unravel_diagram}
\end{figure}

While single-qubit operators are straightforward to encode, the same cannot be said for 2-qubit gates. As these effectively \emph{entangle} states together, we extend the aforementioned unraveling formalism from states to operators. Given a general non-separable $4\times4$ matrix (unitary or not, see below) $\hat{U}$ with elements $u_{\mu\nu}$, $\mu,\nu\in\{1,2,3,4\}$,
\begin{equation}
    \hat{U} = \begin{pmatrix}
    u_{11} & u_{12} & u_{13} & u_{14}\\
    u_{21} & u_{22} & u_{23} & u_{24}\\
    u_{31} & u_{32} & u_{33} & u_{34}\\
    u_{41} & u_{42} & u_{43} & u_{44}
    \end{pmatrix},
    \label{eq:U_block}
\end{equation}
we seek time-dependent $2\times2$ matrices $\hat{S}_1(t),\hat{S}_2(t)$ such that
\begin{equation}
    \hat{U} = {\avginf{\hat{S}_1(t)\otimes\hat{S}_2(t)} \coloneqq \avginf{\hat{U}(t)}}  .
    \label{eq:gate_unravel}
\end{equation}
Fig.~\ref{fig:gate_unravel_diagram} illustrates this structure: the joint action of $\hat{S}_1(t)\otimes\hat{S}_2(t)$ on the two-wavebit system is equivalent to $\hat{S}_1(t)$ and $\hat{S}_2(t)$ acting \emph{independently} on $\ket{{\psi^{(1)}}(t)}$ and $\ket{{\psi^{(2)}}(t)}$ — the operators are locally separable at every instant, and the two-qubit gate is recovered only through the time average at readout.

\begin{figure*}[!t]
\normalsize
\addtocounter{equation}{1}
\begin{align}
    a_1(t) &= \sqrt{u_{11}}s_1(t) + \sqrt{u_{12}}s_2(t) + \sqrt{u_{21}}s_3(t) + \sqrt{u_{22}}s_4(t), \nonumber\\
    b_1(t) &= \sqrt{u_{13}}s_5(t) + \sqrt{u_{14}}s_6(t) + \sqrt{u_{23}}s_7(t) + \sqrt{u_{24}}s_8(t), \nonumber\\
    c_1(t) &= \sqrt{u_{31}}s_9(t) + \sqrt{u_{32}}s_{10}(t) + \sqrt{u_{41}}s_{11}(t) + \sqrt{u_{42}}s_{12}(t), \nonumber\\
    d_1(t) &= \sqrt{u_{33}}s_{13}(t) + \sqrt{u_{34}}s_{14}(t) + \sqrt{u_{43}}s_{15}(t) + \sqrt{u_{44}}s_{16}(t),
    \label{eq:S1_elements}
\end{align}
\addtocounter{equation}{-2}
\hrulefill
\end{figure*}

This is possible by selecting a set of 16 mutually orthonormal signals $\{s_k(t)\}_{k=1}^{16}$ satisfying $\avginf{s_k(t) s_{k'}^*(t)} = \delta_{kk'}$, i.e., orthonormal with respect to the inner product $\langle f,g\rangle\coloneqq\avginf{f^*(t)\,g(t)}$ induced on signals by the time average, where $\delta_{kk'}$ is the Kronecker delta (not to be confused with the two-qubit gate count $\delta$ defined in Sec.~\ref{subsec:sandwich}), and writing the two local operators as $2\times2$ matrices of signals,
\begin{equation}
    \hat{S}_1(t)=\begin{pmatrix}a_1(t) & b_1(t)\\ c_1(t) & d_1(t)\end{pmatrix},\quad
    \hat{S}_2(t)=\begin{pmatrix}a_2(t) & b_2(t)\\ c_2(t) & d_2(t)\end{pmatrix}.
    \label{eq:S_matrices}
\end{equation}
\addtocounter{equation}{1}
The elements of $\hat{S}_1(t)$ are defined in Eq.~\eqref{eq:S1_elements},
and those of $\hat{S}_2(t)$ are constructed analogously from the conjugate signals $\{s_k^*(t)\}$ (see Appendix~\ref{app:gate_expand} for the full expansion; explicit unravelings of common gates, CNOT, CZ, SWAP, are collected in the Supplemental Material~\cite{ref_2_supp}, Sec.~S1). Each surviving time-averaged product is $\smash{\avginf{s_k(t) s_k^*(t)}} = 1$; all cross-terms $\smash{\avginf{s_k(t) s_{k'}^*(t)} = 0}$ for $k\neq k'$ vanish by orthogonality, recovering every element of $\hat{U}$; any branch of the square roots may be used, provided the same one enters $\hat S_1$ and $\hat S_2$.
In this work, the 16 mutually orthonormal signals are built from harmonics of a distinct \emph{base frequency} $\Omega$ (e.g. $s_k(t)=e^{\mathrm{i}k\Omega t}$, $k=1,\ldots,16$), which we elaborate in the following.\\

\textit{Note on non-unitary operations.} This construction never uses the unitarity of $\hat U$: it relies only on the orthonormality of the signals, and therefore unravels \emph{any} $4\times4$ matrix, including non-unitary ones.%

\subsection{Arbitrary unitary circuits}
\label{subsec:arbitrary_circuits}

Any arbitrary circuit can be decomposed into gates belonging to a universal set, typically \(\{\mathrm{Clifford}+T\}\). Above, we showed how to unravel a single two-qubit gate into an analog representation, a process that introduces a base frequency (BF) and its associated harmonics. Single-qubit gates are applied as static matrices (Sec.~\ref{subsec:1qubit_gate}); only the entangling two-qubit gates are unraveled. When extending this construction to circuits composed of multiple gates, a natural question arises: how should these base frequencies be assigned? Since unraveled gates act sequentially on a wavebit, their signals multiply during circuit execution, so that their frequencies add. This introduces a potential problem: the frequency contribution of one gate may be canceled by a combination of frequencies from other gates. \\

To avoid such degeneracies, we require the set of gate-specific BFs, \(\{\Omega_\ell\}\), to be \emph{linearly independent over the rationals}, meaning that no nontrivial integer combination $\sum_\ell k_\ell \Omega_\ell = 0 $ exists for any set of integers \(\{k_\ell\}\). A rigorous formulation of this requirement is provided in Appendix~\ref{app:thm1}. Noticing that the square root of prime numbers are irrational, a practical frequency assignment satisfying this \(\mathbb{Q}\)-independence condition is
\begin{equation}
    \Omega_\ell = \sqrt{p_\ell}\,\omega_\mathrm{ref},
    \label{eq:BF_assign}
\end{equation}
where \(p_\ell\) denotes the \(\ell\)-th prime number and \(\omega_\mathrm{ref}\) is a fixed reference frequency, and $\ell=1,2,\ldots$ numbers the entangling two-qubit gates of the circuit in the order in which they are applied; single-qubit gates are not counted.

\subsection{Example: Sandwich circuit}\label{subsec:sandwich}
\begin{figure*}[t]
\centering
\includegraphics{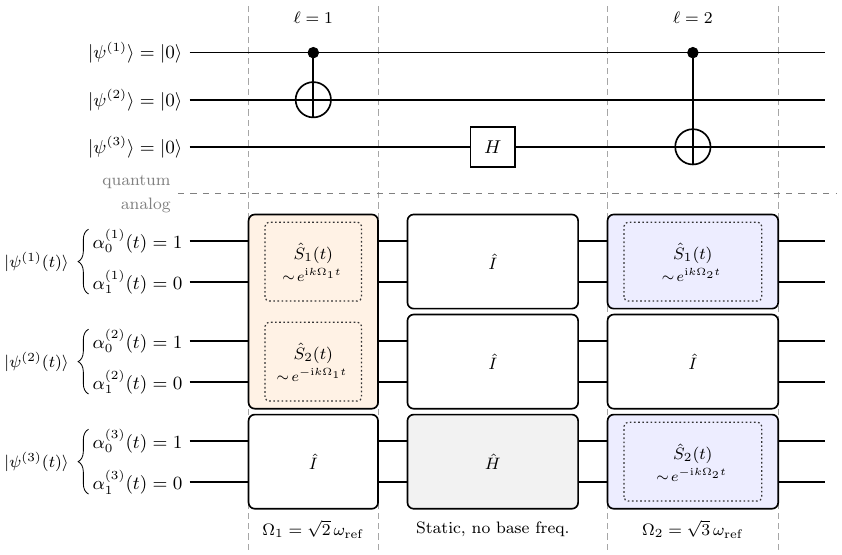}
\caption{\textbf{Quantum and analog representations of the sandwich circuit.} \emph{Top:} Standard quantum circuit ($|{\psi^{(1)}}\rangle$, $|{\psi^{(2)}}\rangle$, $|{\psi^{(3)}}\rangle$) with layer boundaries (dashed lines) and layer labels $\ell$ above. \emph{Bottom:} Analog circuit with six signal channels $[\alpha_0^{(n)},\alpha_1^{(n)}]$ for state $|{\psi^{(n)}}(t)\rangle$, $n=1,2,3$. Entangling layers $\hat{G}_1 = \hat{\mathrm{CNOT}}\otimes \hat{I}$ ($\Omega_1=\sqrt{2}\,\omega_\mathrm{ref}$, orange, acting on $|{\psi^{(1)}}(t)\rangle$ and $|{\psi^{(2)}}(t)\rangle$) and $\hat{G}_2 = P_{1,3}^{-1} (\hat{\mathrm{CNOT}}\otimes\hat{I})P_{1,3}$ ($\Omega_2=\sqrt{3}\,\omega_\mathrm{ref}$, blue, acting on $|{\psi^{(1)}}(t)\rangle$ and $|{\psi^{(3)}}(t)\rangle$) each decompose into local matrices $\hat{S}_1(t)$ and $\hat{S}_2(t)$ whose elements oscillate at harmonics $e^{\pm\mathrm{i}k\Omega_\ell t}$ (dashed inner boxes), with the untouched channel left as $\hat{I}$. The Hadamard step between them, $\hat{H}^{(3)}=\hat{I}\otimes\hat{I}\otimes\hat{H}$, is not an entangling layer: it consists of static single-wavebit matrices only, the Hadamard gate (gray) acting on the channels of $|{\psi^{(3)}}(t)\rangle$ and identities on the other two wavebits, with no oscillating signals. Base frequencies $\Omega_\ell$ appear below the analog section;  two-qubit gate count $\delta=2$.}
\label{fig:cnot_cascade}
\end{figure*}
As an example for the following, we now construct the analog circuit corresponding to a specific quantum circuit: the sandwich circuit, in which a Hadamard gate is \emph{sandwiched} between two controlled-NOT gates (upper panel of Fig.~\ref{fig:cnot_cascade}). The corresponding unitary is written as a product of two \emph{entangling layers}, one per two-qubit gate, with the Hadamard gate applied between them, all acting on the full three-qubit state,
\begin{equation}
    \begin{aligned}
        \hat{G}_\mathrm{sand}
        &= \hat{G}_2\hat{H}^{(3)}\hat{G}_1
    \end{aligned}
    \label{eq:CNOT_cascade}
\end{equation}
where,
\begin{equation}
\begin{aligned}
    \hat{G}_1 &=  {\hat{\mathrm{CNOT}} \otimes \hat{I}}, \\
    {\hat{G}_2} &=  P_{1,3}^{-1}(\hat{\mathrm{CNOT}}\otimes\hat{I})P_{1,3},
\end{aligned}
\label{eq:CNOT_cascade_layers}
\end{equation}
with $\hat{\mathrm{CNOT}}$ the $4\times4$ controlled-NOT gate acting on the first two tensor factors, $\hat{H}^{(3)}=\hat{I}\otimes\hat{I}\otimes\hat{H}$ the Hadamard gate acting on the third qubit, and $P_{1,3}$ the permutation operator reordering the tensor factors.\\%

To unravel this circuit, each layer $\hat{G}_\ell$ is attributed a time-dependent matrix $\hat{G}_\ell(t)$ whose time-average recovers the original gate, $\avginf{\hat{G}_\ell(t)} = \hat{G}_\ell$, with a unique base frequency $\Omega_\ell$ assigned to each layer. Recalling that two-qubit gates can be unraveled into two $2\times2$ matrices $\hat{S}_1(t)$ and $\hat{S}_2(t)$ (Eq.~\eqref{eq:gate_unravel}), the two layers of the sandwich circuit are unraveled as
\begin{equation}
\begin{aligned}
    \hat{G}_1(t) &= \hat{U}_{1}(t)\otimes \hat{I}, \\
    {\hat{G}_2(t)} &=  P_{1,3}^{-1}\left( {\hat{U}_{2}(t)}\otimes\hat{I}\right)P_{1,3},
\end{aligned}
\label{eq:active_layers}
\end{equation}
where each CNOT gate is unraveled into a control matrix $\hat{S}_1(t)$ and a target matrix $\hat{S}_2(t)$,
\begin{equation}
    \hat{U}_{\ell}(t) = \begin{pmatrix}e^{\mathrm{i}\Omega_\ell t} & 0\\0 & e^{2\mathrm{i}\Omega_\ell t}\end{pmatrix}
    \otimes\begin{pmatrix}e^{-\mathrm{i}\Omega_\ell t} & e^{-2\mathrm{i}\Omega_\ell t}\\e^{-2\mathrm{i}\Omega_\ell t} & e^{-\mathrm{i}\Omega_\ell t}\end{pmatrix},
    \label{eq:CNOT_unravel}
\end{equation}
 with $\ell\in\{1,2\}$: the control wavebit selects between the two harmonics $e^{\mathrm{i}\Omega_\ell t}$ and $e^{2\mathrm{i}\Omega_\ell t}$, and only the matching harmonic on the target wavebit survives the time average, giving the identity for control $\ket{0}$ and the bit flip for control $\ket{1}$.\\

 Each CNOT gate thus uses only two orthonormal harmonics of its base frequency $\Omega_\ell$ (following Eq.~\eqref{eq:BF_assign}, $\Omega_1=\sqrt{2}\,\omega_\mathrm{ref}$ and $\Omega_2=\sqrt{3}\,\omega_\mathrm{ref}$; any other $\mathbb{Q}$-independent pair would do). The Hadamard gate, in contrast, is not an entangling layer: it is applied as a static matrix to the two channels of $\ket{{\psi^{(3)}}(t)}$ (Sec.~\ref{subsec:1qubit_gate}), while $\ket{{\psi^{(1)}}(t)}$ and $\ket{{\psi^{(2)}}(t)}$ are left untouched. It requires no oscillating signal and is assigned no base frequency; in particular, it is not passed through the construction of Eq.~\eqref{eq:S1_elements}, which for $\Omega=0$ would collapse all signals $s_k(t)$ to unity and lose their orthonormality. The $\mathbb{Q}$-independence of $\{\Omega_1,\Omega_2\}$ ensures that the time averages of the two layers decouple, each recovering its respective CNOT independently, so that $\avginf{\hat{G}_2(t)\hat{H}^{(3)}\hat{G}_1(t)}=\hat{G}_2\hat{H}^{(3)}\hat{G}_1=\hat{G}_\mathrm{sand}$ (Theorem~\ref{thm:freq_indep}). The lower panel of Fig.~\ref{fig:cnot_cascade} illustrates the resulting analog circuit, corresponding to the quantum one above, composed of six signal channels $\{\alpha_0^{(n)}(t),\alpha_1^{(n)}(t)\},\quad n = 1,2,3$ (two signals per qubit).

Although the sandwich circuit serves as a concrete example for circuit unraveling, the procedure naturally extends to arbitrary $N$-qubit quantum circuits, as detailed in Appendix~\ref{app:circuit_univ}.

Finally, as a convention for the remainder, an \emph{entangling layer}, or simply \emph{layer}, contains exactly one two-qubit gate, and single-qubit gates are applied between layers as static matrices. We define the \emph{two-qubit gate count} $\delta$ as the number of entangling layers in a circuit, i.e., of two-qubit gates, with layer index $\ell$ coinciding with its assigned base frequency $\Omega_\ell$.

\subsection{Finite-time error bound}
\label{subsec:error}
Before moving on to experimental and simulation results, we briefly discuss the error incurred when the idealized infinite-time average is replaced by a finite averaging window $T$.  The resulting error between the finite-$T$ and infinite-$T$ averages is bounded analytically. Let us define a sum that quantifies the convergence of the time average:
\begin{equation}
    S = \sum_{\substack{\mathbf{k}\neq\mathbf{0}}}
    \frac{\|\hat{A}_{1,k_1}\cdots\hat{A}_{\delta,k_\delta}\|}
    {\left|\sum_\ell k_\ell\Omega_\ell\right|},
    \label{eq:S_sum}
\end{equation}
where $A_\ell(t)$ is the time-varying operator of layer $\ell$, with fundamental frequency $\Omega_\ell$ (constant single-qubit gates can be absorbed into it), $\hat{A}_{\ell,k}$ is its $k$-th Fourier component, and $\|\cdot\|$ is any sub-multiplicative matrix norm.  If $S$ converges, then
\begin{equation}
    \xi := \left\|\avginf{A_1\cdots A_\delta} - \avgT{A_1\cdots A_\delta}\right\| \leq \frac{2S}{T}.
    \label{eq:err_bound}
\end{equation}

\noindent (Proof in Appendix~\ref{app:thm2}).  The bound gives a prescribed
averaging time $T > 2S/\gamma$ to achieve error $\xi < \gamma$.  Using bandlimited signals for the matrix decomposition ensures convergence of $S$.

\section{Analog Hardware Implementation}
\label{sec:hardware}

\subsection{Architecture}

\begin{figure}[t]
\centering
\includegraphics{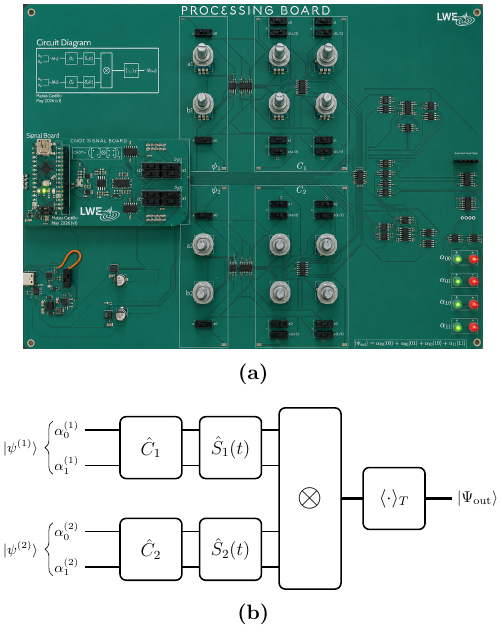}
\caption{\textbf{Emulator architecture and physical implementation.} \textbf{(a)} Fully assembled mixed-signal prototype implementing the operations in~(b). The Signal Board (left) generates the BF signal at $1\,\mathrm{kHz}$ and its third harmonic at $3\,\mathrm{kHz}$. The Processing Board (right) implements the analog tensor product and time-averaging filter described in the Supplemental Material~\cite{ref_2_supp}, Sec.~S2. \textbf{(b)} Main operations implemented by the emulator, as realized on the hardware in~(a). Signal flow from the two separated input states $|{\psi^{(1)}}\rangle$ and $|{\psi^{(2)}}\rangle$ (with amplitudes $\alpha_0^{(1)},\alpha_1^{(1)}$ and $\alpha_0^{(2)},\alpha_1^{(2)}$ respectively) through the static encoding matrices $\hat{C}_1,\hat{C}_2$ (single-qubit gates), the time-varying signal matrices $\hat{S}_1(t),\hat{S}_2(t)$ (BF carriers), the analog tensor product ($\otimes$), and the time-averaging filter ($\langle\cdot\rangle_T$) to yield the output state coefficients. }
\label{fig:hardware}
\end{figure}

To validate the theoretical framework, we constructed a mixed-signal printed circuit board (PCB) system capable of emulating 2-qubit unitary circuits as shown in Fig.~\ref{fig:hardware}a. The hardware comprises two boards: a Signal Board that generates the BF signals through Direct Digital Synthesis (DDS), and a Processing Board that implements the analog tensor product and time-averaging filter.  The two boards are connected via header pins.  The emulator is controlled by a microcontroller that programs the DDS chips, and by potentiometers that the user utilizes to set the gain of a set of VGAs.  A complete description of the hardware design, assembly, and calibration is provided in the Supplemental Material~\cite{ref_2_supp}, Sec.~S2.

\textit{Operating parameters.}  In our experiments, $\Omega_1/(2\pi) = 1\,\mathrm{kHz}$ is the single base frequency driving the CNOT layer, and $3\Omega_1/(2\pi) = 3\,\mathrm{kHz}$ is the second orthonormal carrier (Eq.~\eqref{eq:CNOT_unravel}).  The 3rd harmonic is chosen, rather than the 2nd, to increase the spectral separation between the two carriers and reduce cross-talk.  Since the prototype implements a single entangling layer, only one BF is required; the $\mathbb{Q}$-independence condition, which ensures independent time averages across multiple layers, is not needed here.
The four output coefficients are produced \emph{in parallel} as DC voltages behind four second-order Sallen--Key averaging filters with $f_c\approx2\,\mathrm{Hz}$ (Supplemental Material~\cite{ref_2_supp}, Sec.~S2), so a complete two-wavebit output state is available on a sub-second timescale set by the filter settling, independent of which circuit is programmed — with the exception of the phase-kickback input, where a slow multiplier artifact required longer oscilloscope averaging (Supplemental Material~\cite{ref_2_supp}, Sec.~S3).

 Finally, for simplicity, the current prototype operates with real-valued BF signals and real output coefficients.

\subsection{Experimental results}
\label{subsec:experimental_results}

\begin{figure*}[htb]
\centering
\includegraphics{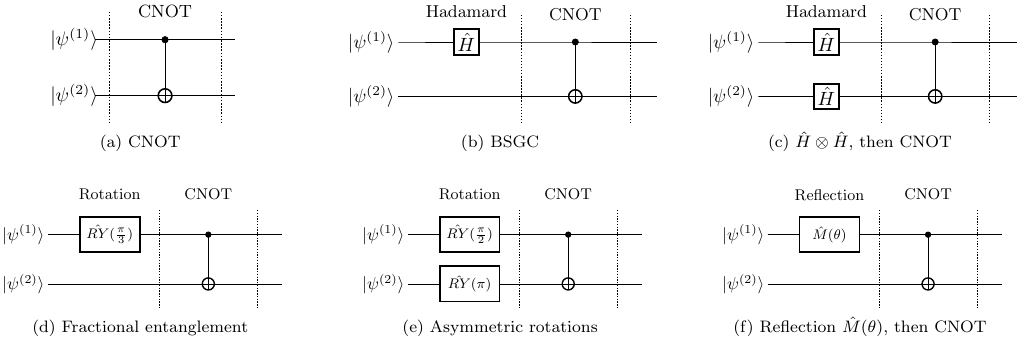}
\caption{\textbf{Experimental circuits tested on the analog prototype.} Each panel corresponds to one circuit family in Table~\ref{tab:fidelities}. The CNOT target qubit is drawn with the mixer symbol ($+$ circle, implemented in the analog hardware as a ring-multiplier element). (a)~Bare CNOT, used for both the computational-basis and phase-kickback ($\ket{+}\ket{-}$) inputs. (b)~Bell State Generation Circuit (BSGC): Hadamard on $|{\psi^{(1)}}\rangle$ followed by CNOT. (c)~Hadamard on both wavebits followed by CNOT. (d)–(e)~$R_Y$ rotations at $\pi/3$ and $\pi/2$ (with $\pi$ on $|{\psi^{(2)}}\rangle$) before CNOT, probing fractional and asymmetric entanglement. (f)~Bloch-sphere reflection $\hat{M}(\theta)$ at $\theta=53.1^\circ$ before CNOT. Dashed vertical lines mark the gate boundaries (state preparation $\mid$ gate application $\mid$ readout).}
\label{fig:circuits}
\end{figure*}

We tested six distinct 2-qubit circuits across all four computational basis input states; these are shown in Fig.~\ref{fig:circuits}. Performance is quantified by the state fidelity
\begin{equation}
    \mathcal{F} = \bigl|\langle\Psi_\mathrm{meas}|\Psi_\mathrm{target}\rangle\bigr|^2,
    \label{eq:exp_fidelity}
\end{equation}
computed from the DC output voltages, rescaled by the machine unit (Supplemental Material~\cite{ref_2_supp}, Sec.~S3), against the ideal target state for each input. Table~\ref{tab:fidelities} summarizes the state fidelities and amplitude errors for all tested circuits (two tests were performed on the same CNOT circuit). Raw circuit output results are in Supplemental Material~\cite{ref_2_supp}, Sec.~S3.

\begin{table}[h]
\centering
\caption{\textbf{State fidelities and amplitude errors for tested circuits.}
$\mathcal{F}=|\langle\Psi_\mathrm{meas}|\Psi_\mathrm{target}\rangle|^2$ averaged over the inputs tested (the four computational basis states; $\ket{+}\ket{-}$ for phase kickback). $\mathcal{F}_\mathrm{norm}$: same with $\Psi_\mathrm{meas}$ normalized to unit norm, which removes the dissipative losses of the electronic circuit.
$|\Delta\psi|_\mathrm{max}$: maximum single-amplitude absolute deviation $\max_{b,\,\mathrm{input}}|\alpha_b^\mathrm{meas}-\alpha_b^\mathrm{target}|$ over all basis states and all inputs tested.}
\label{tab:fidelities}
\begin{tabular}{lccc}
\toprule
Circuit & $|\Delta\psi|_\mathrm{max}$ & $\mathcal{F}$ & $\mathcal{F}_\mathrm{norm}$ \\
\midrule
CNOT (computational basis)              & 0.026 & 0.96 & $>0.999$ \\
CNOT (phase kickback: $\ket{+}\ket{-}$) & 0.047 & 0.94 & 0.995 \\
Bell State Generation (BSGC)            & 0.021 & 0.96 & $>0.999$ \\
$\hat{H}\otimes\hat{H}$ then CNOT      & 0.059 & 0.92 & 0.995 \\
Fractional entanglement ($R_Y(\pi/3)$)  & 0.017 & 0.98 & $>0.999$ \\
Asymmetric tensor product               & 0.023 & 0.95 & $>0.999$ \\
Reflection ($\hat{M}(53.1^\circ)$)     & 0.018 & 0.97 & $>0.999$ \\
\bottomrule
\end{tabular}
\end{table}

Across all configurations, the maximum single-amplitude error is $|\Delta\psi|_\mathrm{max}\le0.059$, dominated by manual potentiometer-setting uncertainty and residual filter ripple rather than by the encoding itself. This result is confirmed through the high fidelity achieved across all circuits. The Bell State Generation Circuit (BSGC) --- the most demanding test, requiring the emulator to output the amplitudes of a maximally nonseparable state --- reached $\mathcal{F}=0.96$, demonstrating that classical analog hardware can faithfully reproduce the amplitude structure of entangled states within this encoding. The fidelity can be further increased by accounting for the dissipative losses in the electronic circuit, which reduce the norm of the measured states (Supplemental Material~\cite{ref_2_supp}, Sec.~S3): normalizing them yields $\mathcal{F}_\mathrm{norm}\ge0.995$ for all circuits.\\

\section{Numerical Simulations}
\label{sec:simulations}

The numerical results in this section are digital simulations \emph{of the wavebit emulator itself} — the dynamic signals, layer operators, and time averages are computed explicitly and compared against exact statevectors (Qiskit) — and serve to validate the framework beyond the two-wavebit scale of the hardware prototype.

\subsection{Averaging Window and Fidelity}
\label{subsec:sim_states}

\begin{figure}[t]
\centering
\includegraphics[width=\columnwidth]{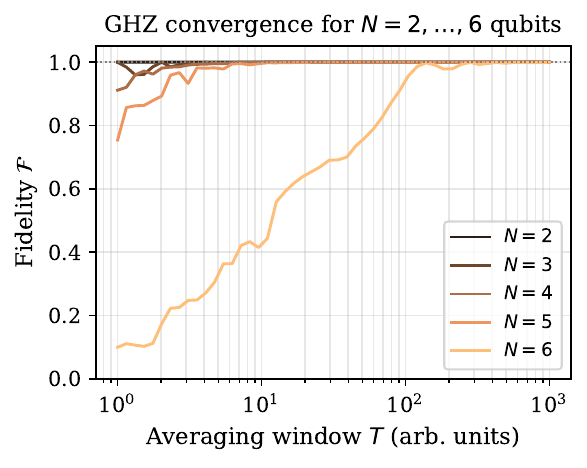}
\caption{\textbf{Multi-qubit GHZ-state convergence.} Fidelity of the time-averaged state with the target GHZ state $\ket{\mathrm{GHZ}_N}=(\ket{0}^{\otimes N}+\ket{1}^{\otimes N})/\sqrt{2}$ as a function of averaging window $T$ for $N=2,3,4,5,6$ qubits (copper colormap, dark to light). All cases converge to unit fidelity; the convergence window grows with $N$ due to the increased number of NSC frequency components — the $N=6$ curve requires $T\sim10^3$ arb.\ units. Frequencies chosen as $\Omega_n=\sqrt{p_n}\,\omega_\mathrm{ref}$ for distinct primes $p_n$.}
\label{fig:multiqubit}
\end{figure}

Here, we show that resolving larger states requires longer averaging windows. To do so, we simulate GHZ states $\ket{\mathrm{GHZ}_N}=(\ket{0}^{\otimes N}+\ket{1}^{\otimes N})/\sqrt{2}$ using a systematic procedure that generates the dynamic state corresponding to an $N$-qubit state \emph{directly} (see Appendix~\ref{app:gate_unravel}). Fidelity $\mathcal{F}=|\braket{\Psi_\mathrm{avg}}{\Psi_\mathrm{target}}|^2$ is computed from the numerically time-averaged state. Fig.~\ref{fig:multiqubit} shows that all five GHZ states ($N=2,\ldots,6$) converge to unit fidelity as $T\to\infty$. The convergence window grows with $N$ because higher-qubit constructions involve more NSC frequency components, increasing the sum $S$ in the error bound $\xi\le 2S/T$. For $N=6$ the required window extends to $T\sim10^3$ arb.\ units, consistent with the larger number of near-resonant combination frequencies.

\subsection{Active gate simulation and circuit benchmarks}
\label{subsec:sim_circuits}

\begin{figure*}[t]
\centering
\includegraphics{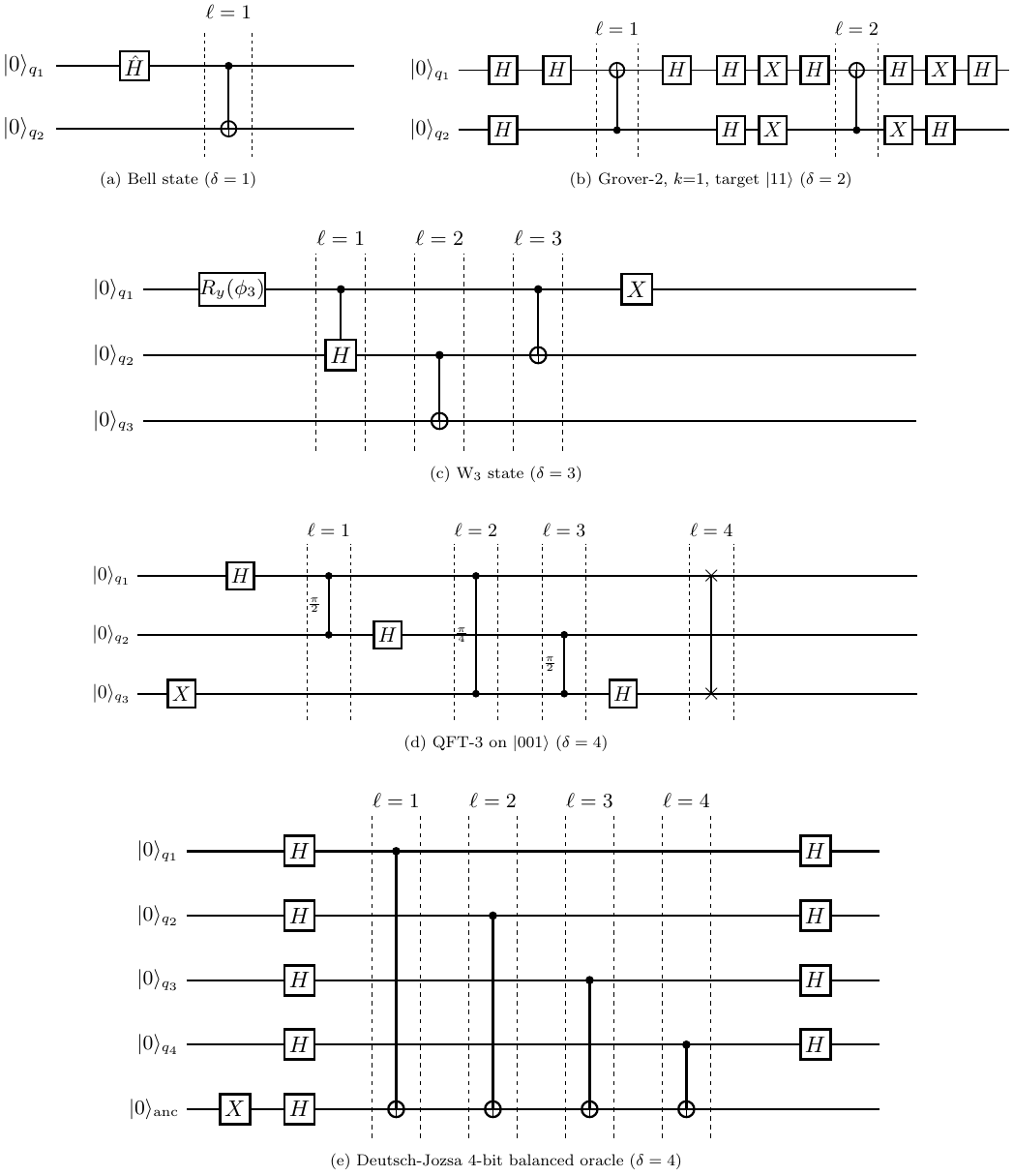}
\caption{\textbf{Benchmark circuits and their layer decompositions.} Circuits are ordered by increasing two-qubit gate count $\delta$ (= number of base frequencies required). Each two-qubit gate occupies one entangling layer (dashed boundaries, labeled $\ell=1,2,\ldots$), driven at a $\mathbb{Q}$-independent base frequency $\Omega_\ell=\sqrt{p_\ell}\,\omega_\mathrm{ref}$. Single-qubit gates (rectangles) are static matrices applied between layers, requiring no oscillating signals. In (b), the oracle ($U_f$, CZ realized as $\hat{H}\cdot\text{CX}\cdot\hat{H}$) and diffusion ($D$) each contain exactly one CNOT layer ($\ell=1$ and $\ell=2$ respectively). In (c), the controlled-$H$ gate ($\ell=1$) and two CNOT gates ($\ell=2$, $\ell=3$) prepare the W$_3$ state from $|000\rangle$. CP gates in (d) use a two-filled-circle symbol; the SWAP gate is shown as crossed wires ($\times$). In (e), all four CNOT gates ($\ell=1$ to $\ell=4$) each target the same ancilla qubit from a distinct input-qubit control. The DJ constant oracle (not shown) is identical to the balanced circuit of (e) with the four CNOT gates removed, yielding $\delta=0$ and exact output for any averaging window.}
\label{fig:sim_circuits}
\end{figure*}

We simulate six circuits using the layer decomposition method as discussed in Sec.~\ref{subsec:arbitrary_circuits}.  Each 2-qubit gate forms one layer $\hat{G}_\ell$ driven at its own base frequency $\Omega_\ell = \sqrt{p_\ell}\,\omega_\mathrm{ref}, \quad p_\ell\text{ the }\ell\text{-th prime}$. Single-qubit gates are applied as static (un-oscillating) matrices, and every two-qubit gate, the CNOT included, is unraveled with the 16 harmonics of the general construction of Eq.~\eqref{eq:S1_elements}, with $\omega_\mathrm{ref}=2\pi\times1\,$kHz. The imposed independence of the frequencies $\Omega_\ell$ ensures that the time averages of all entangling layers decouple, each recovering the effective quantum operation. The averaging window spans twenty periods of the slowest combination frequency allowed by this construction, $T=20\times2\pi/\omega^*_\delta$ with $\omega^*_\delta=\min_{\mathbf{k}\neq\mathbf{0}}|\sum_\ell k_\ell\Omega_\ell|$ ($|k_\ell|\le15$), a conservative choice, since not all of these combinations carry weight in a given circuit, and the time average is evaluated as the mean over samples taken at $F_s=100\,$kHz (Supplemental Material~\cite{ref_2_supp}, Secs.~S4 and~S5). Fidelity is computed against the exact Qiskit statevector; see Supplemental Material~\cite{ref_2_supp}, Sec.~S6 for details regarding the simulation. \\

\paragraph{Bell state $(\delta=1)$.}
The Bell-state generation circuit (Fig.~\ref{fig:sim_circuits}a) applies a Hadamard gate on $\ket{\psi^{(1)}}$ followed by a single CNOT gate, constituting one layer $\hat{G}_1$ at $\Omega_1=\sqrt{2}\,\omega_\mathrm{ref}$. With a single layer there are no inter-layer cross terms, and the time average is exact to machine precision whenever $T$ spans a whole number of periods $2\pi/\Omega_1$, as it does here. This baseline confirms that the single-gate unraveling and Qiskit little-endian embedding are correct.\\

\paragraph{2-qubit Grover search $(\delta=2)$.}
Fig.~\ref{fig:sim_circuits}b shows Grover's algorithm on $n=2$ qubits targeting the state $\ket{11}$. The oracle $U_f$ (a CZ gate, realized as $\hat{H}\cdot\text{CX}\cdot\hat{H}$) and the diffusion operator $D=2\ket{s}\bra{s}-\hat{I}$ each contribute one CNOT layer; together they constitute two layers $\hat{G}_1$ and $\hat{G}_2$. For $n=2$, the angle $\theta=\arcsin(2^{-n/2})=\pi/6$ is such that a single Grover iteration, a rotation by $2\theta$, takes the state from angle $\theta$ to $3\theta=\pi/2$, i.e., exactly onto the target, giving $P(\ket{11})=1$.\\

\paragraph{3-qubit $\mathrm{W}_3$ state $(\delta=3)$.}
The $\mathrm{W}_3$ state $\ket{W_3}=\frac{1}{\sqrt{3}}(\ket{001}+\ket{010}+\ket{100})$ encodes exactly one excitation shared equally among three qubits and represents the W-class of tripartite entanglement, inequivalent to GHZ-type entanglement under local operations and classical communication. The preparation circuit (Fig.~\ref{fig:sim_circuits}c) applies $R_y(\phi_3)$ with $\phi_3=2\arccos(1/\sqrt{3})$ to qubit $q_1$, followed by a controlled-$H$ gate ($q_1\to q_2$), two CNOT gates ($q_2\to q_3$ then $q_1\to q_2$), and a final $\hat{X}$ on $q_1$. After transpilation to the CX basis, three layers $\hat{G}_1,\hat{G}_2,\hat{G}_3$ emerge at frequencies $\sqrt{2},\sqrt{3},\sqrt{5}$ (in units of $\omega_\mathrm{ref}$).\\

\paragraph{3-qubit QFT $(\delta=4)$.}
The quantum Fourier transform on $\ket{001}$ (Fig.~\ref{fig:sim_circuits}d) transpiles to three controlled-phase ($\hat{CP}$) gates and one SWAP gate, giving $\delta=4$ layers at frequencies $\sqrt{2},\ \sqrt{3},\ \sqrt{5},\ \sqrt{7}$ (in units of $\omega_\mathrm{ref}$). The QFT is a stringent phase test because the structured phase relationships between its $\hat{CP}$ gates require all four layer averages to cooperate.\\

\begin{figure*}[t]
\centering
\includegraphics[width=\textwidth]{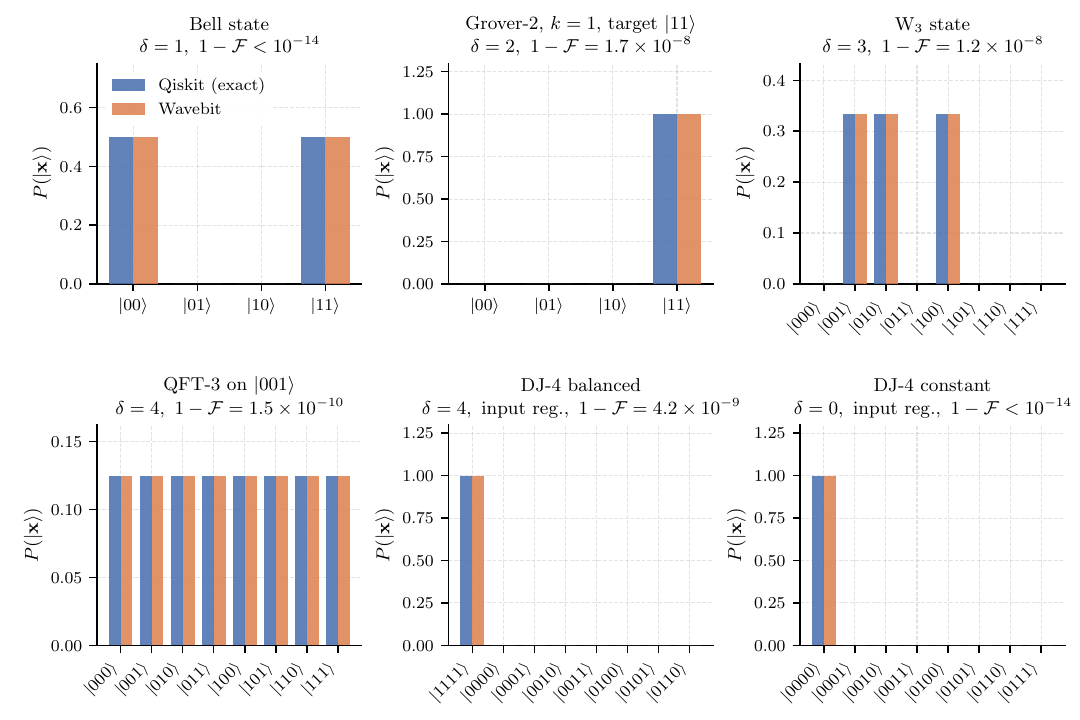}
\caption{\textbf{Output probability distributions: Qiskit (exact) vs.\ wavebit (simulated).} Each panel shows computational-basis probabilities $P(|\mathbf{x}\rangle)=|\langle\mathbf{x}|\Psi\rangle|^2$ for the exact Qiskit statevector (blue) and the time-averaged wavebit simulation (orange), with the settings of Table~\ref{tab:sim_results}. Panel titles quote the infidelity $1-\mathcal{F}$ and entangling two-qubit gate count $\delta$. \textit{Top row (left to right):} Bell state ($\delta=1$) — equal weight on $\ket{00}$ and $\ket{11}$ at machine precision; Grover-2 ($\delta=2$) — $P(\ket{11})\to1$; $\mathrm{W}_3$ state ($\delta=3$) — equal weight on the three single-excitation states. \textit{Bottom row:} QFT-3 on $\ket{001}$ ($\delta=4$) — flat 8-state spectrum; DJ-4 balanced oracle ($\delta=4$, input register) — $P(\ket{0000})=0$ correctly identifies the balanced function, all probability being concentrated on $\ket{1111}$ for this parity oracle; DJ-4 constant oracle ($\delta=0$, input register) — $P(\ket{0000})=1$ identifies the constant function exactly, with no averaging required.}
\label{fig:sim_probs}
\end{figure*}

\paragraph{Deutsch-Jozsa --- balanced oracle $(\delta=4)$.}
The circuit in Fig.~\ref{fig:sim_circuits}e decides whether $f:\{0,1\}^4\to\{0,1\}$ is constant or balanced. The balanced oracle $f=x_0\oplus x_1\oplus x_2\oplus x_3$ consists of four CNOT gates, each connecting a distinct input qubit to a shared ancilla — four layers $\hat{G}_1,\ldots,\hat{G}_4$ at $\Omega_\ell=\sqrt{p_\ell}\,\omega_\mathrm{ref}$. The decision rule, evaluated on the four-qubit input register, is $P(\ket{0000})=0$ (balanced) vs.\ $P(\ket{0000})=1$ (constant); for this parity oracle, the balanced output in fact concentrates all probability on $\ket{1111}$ (Fig.~\ref{fig:sim_probs}).\\%

\paragraph{Deutsch-Jozsa --- constant oracle $(\delta=0)$.}
The constant oracle ($f\equiv0$) contains no CNOT gates, removing all entangling layers ($\delta=0$). Without any oscillating signals, the wavebit state is time-independent and the time average is exact for any window; the emulator returns $P(\ket{0000})=1$, correctly identifying the constant function.\\

All results are collected in Table~\ref{tab:sim_results} and Fig.~\ref{fig:sim_probs} shows the full output probability distributions. Across the benchmarks computed here ($\delta\le4$), the amplitude error decreases as $\mathcal{O}(1/T)$, and hence the infidelity as $\mathcal{O}(1/T^2)$, in agreement with the bound of Sec.~\ref{subsec:error}, with a $\delta$-dependent prefactor: the $\delta=4$ circuits (QFT-3 and DJ-4) converge most slowly because their four layer frequencies generate the densest set of near-resonant combination frequencies (see Supplemental Material~\cite{ref_2_supp}, Sec.~S4 for the convergence curves and beat-frequency analysis). The smallest combination frequency $\omega^*_\delta$ shrinks much faster than the spacing of the carriers. Even with the two-harmonic unraveling of Eq.~\eqref{eq:CNOT_unravel}, each base frequency enters a readout product with coefficient $-1$, $0$ or $+1$, so that $3^\delta$ combination frequencies are packed into a range of width $\sim\delta\sqrt{p_\delta}\,\omega_\mathrm{ref}$; the smallest nonzero one is therefore generically exponentially small in $\delta$ (numerically, $\approx0.09\,\omega_\mathrm{ref}$ at $\delta=4$ and $5\times10^{-4}\,\omega_\mathrm{ref}$ at $\delta=10$). The 16-harmonic construction of Eq.~\eqref{eq:S1_elements} only packs them more densely, giving $\omega^*_4\approx1.3\times10^{-4}\,\omega_\mathrm{ref}$ for the circuits of Table~\ref{tab:sim_results}. In the worst case, the averaging time grows exponentially with the number of two-qubit gates. Incidentally, the time average acts as a Feynman path sum, in which each frequency combination labels one path through the circuit and only frequency-matched paths survive the average. Like path-sum simulation, it is cheap in memory ($2N$ channels) but exponential in gate count, whereas tensor-network contraction is exponential only in the treewidth of the circuit~\cite{markov_simulating_2008}.

In the physical wavebit system, no sampling is involved: the analog integrator measures $\langle\alpha_\mathbf{b}(t)\rangle_T$ directly in continuous time. The growth of the required window $T$ with the number of two-qubit gates is, however, intrinsic, and the achievable accuracy is ultimately capped by hardware noise and the dynamic range of the analog signal chain. Together with the readout stage, this convergence cost is where the exponential difficulty of classical emulation reappears.

\begin{table}[t]
\centering
\caption{\textbf{Wavebit circuit simulation results.} $N$: number of qubits. $\delta$: entangling two-qubit gate count (= base frequencies required). $|\Delta\psi|_{\max}$: maximum single-amplitude absolute error. $\Delta\phi_{\max}$: maximum single-amplitude phase error (rad). $1-\mathcal{F}$: infidelity vs.\ exact Qiskit statevector. $T$: averaging window, twenty periods of $\omega^*_\delta$ ($\omega_\mathrm{ref}=2\pi\times1\,$kHz, $F_s=100\,$kHz); the constant oracle ($\delta=0$) is exact for any window.}
\label{tab:sim_results}
\begin{ruledtabular}
\setlength{\tabcolsep}{1pt}
\begin{tabular}{lcccccc}
Circuit & $N$ & $\delta$ & $T$ (s) & $|\Delta\psi|_{\max}$ & $\Delta\phi_{\max}$ & $1-\mathcal{F}$ \\
\hline
Bell state           & 2 & 1 & $0.014$ & $<\!10^{-14}$      & $<\!10^{-14}$      & $<\!10^{-14}$ \\
Grover-2, $k{=}1$    & 2 & 2 & $0.62$  & $2.0\times10^{-4}$ & $2.0\times10^{-4}$ & $1.7\times10^{-8}$ \\
W$_3$ state          & 3 & 3 & $11.7$  & $8.2\times10^{-5}$ & $1.3\times10^{-4}$ & $1.2\times10^{-8}$ \\
QFT-3 on $\ket{001}$ & 3 & 4 & $150$   & $7.2\times10^{-6}$ & $1.5\times10^{-5}$ & $1.5\times10^{-10}$ \\
DJ (balanced)        & 5 & 4 & $150$   & $1.7\times10^{-5}$ & $1.7\times10^{-5}$ & $4.2\times10^{-9}$ \\
DJ (constant)        & 5 & 0 & ---     & $<\!10^{-14}$      & $<\!10^{-14}$      & $<\!10^{-14}$ \\
\end{tabular}
\end{ruledtabular}
\end{table}

\section{Discussion and Outlook}
\label{sec:discussion}

\subsection{Limitations}
\label{subsec:limitations}
\begin{figure*}[t]
\centering
\includegraphics{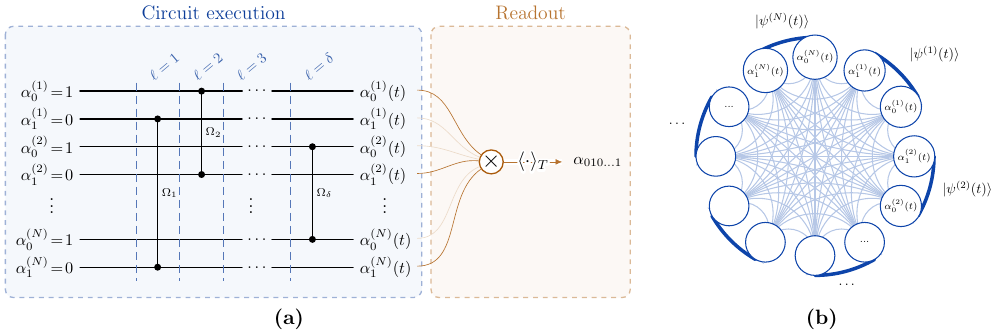}
\caption{\textbf{Proposed $N$-wavebit system and readout chain.}
  \textbf{(a)}~\textit{Circuit execution stage}: the static initial coefficients $\alpha_{b_n}^{(n)}(t)=\alpha_{b_n}^{(n)}$ are injected into the wavebit network, where coupling through auxiliary base frequencies $\Omega_\ell$ drives them into time-dependent coefficients $\alpha_{b_n}^{(n)}(t)$ that encode the circuit evolution. \textit{Readout stage}: for a single bit-string $\mathbf{b}$, the corresponding dynamic coefficients are selected and fed into an $N$-way analog product $\prod_n \alpha_{b_n}^{(n)}(t)$, then time-averaged ($\langle\cdot\rangle_T$) to extract the static coefficient $\alpha_{b_1 b_2\cdots b_N}$. Full readout of all $2^N$ basis coefficients requires $2^N$ such operations. \textbf{(b)}~All-to-all connected array of $N$ wavebits $|{\psi^{(n)}}(t)\rangle$,
  each encoded by two dynamic coefficients $\alpha_{b_n}^{(n)}(t)$ coupled through NSCs; metamaterial arrays of active resonators are a natural physical substrate.}
\label{fig:readout}
\end{figure*}
\paragraph{Readout stage and $2^N$ computational complexity.}
During the \emph{circuit execution stage}, BFs drive the $N$-wavebit state through successive entangling layers, each implemented as a separable time-varying operator whose time average enacts the desired entangling unitary. We have seen that, for a given gate count, the resource cost is irrespective of system size $N$. Generally, it is polynomial in the two-qubit gate count $\delta$ ($N_\mathrm{BF} = \mathcal{O}(\delta)$). The \emph{readout stage}, however, exposes a fundamental asymmetry.

To extract a single static output coefficient $\alpha_{b_1 b_2\cdots b_N}$ from the emulator, one forms the multi-wavebit product
\begin{equation}
    \alpha_{b_1 b_2\cdots b_N}(t)
    = \prod_{n=1}^{N} \alpha_{b_n}^{(n)}(t),
    \label{eq:readout_product}
\end{equation}
and applies the time-averaging filter to the result.  In hardware, this requires an $N$-way analog multiplication (a 2-way multiplier suffices for the current 2-wavebit prototype), followed by a single low-pass filter stage. Fig.~\ref{fig:readout}(a) illustrates a systematic analog implementation of this readout chain.

Recovering the \emph{complete} output state demands repeating this procedure for all $2^N$ computational basis strings $b_1 b_2\cdots b_N$, i.e., $2^N$ independent $N$-way multiplications and time-averages.  The readout hardware complexity is therefore $\mathcal{O}(2^N)$: exponential in qubit count, regardless of the number of gates. This exponential cost at readout is the primary scalability bottleneck of the wavebit architecture.  It does not arise from the circuit execution itself (which remains polynomial in the number of two-qubit gates) but from the need to fully characterize the $2^N$-dimensional output state. In special cases the cost can be reduced: for algorithms whose ideal output is a single computational basis state, a sequential implementation of the readout stage could evaluate coefficients one at a time and terminate as soon as the unit-amplitude coefficient is identified.%

Furthermore, note that the emulator is deterministic: it outputs the complex amplitudes themselves, including their phases, rather than measurement samples, and Born probabilities follow as $|\alpha_{b_1\cdots b_N}|^2$. These cannot be obtained by squaring the signals before averaging, since $\avginf{|\alpha_{b_1\cdots b_N}(t)|^2}$ also contains the power of all oscillating components. Born probabilities of single-qubit measurements therefore inherit the exponential readout cost, since each sums $2^{N-1}$ squared amplitudes.

The readout cost of Born probabilities can, however, be moved from the readout hardware to the averaging time. Running two copies A and B of the register, whose base frequencies (including any offset $\Omega_0$ of the initial-state encoding, Sec.~\ref{subsec:circuits}) are jointly linearly independent over $\mathbb{Q}$, any product observable $\hat O=\hat O_1\otimes\cdots\otimes\hat O_N$ follows from a single readout chain,
\begin{equation}
    \bra{\Psi}\hat O\ket{\Psi}=\avginf{\prod_{n=1}^{N}\bra{\psi^{(n)}_B(t)}\hat O_n\ket{\psi^{(n)}_A(t)}},
    \label{eq:two_copy}
\end{equation}
where the bra contains the complex-conjugated signals of copy B. Since each copy is a product state at every instant, the sum over the $2^N$ basis strings factorizes over wavebits before averaging, and by Theorem~\ref{thm:freq_indep} only the product of the two DC components survives. Choosing $\hat O_n=\ket{b_n}\bra{b_n}$ for $n\le k$ and $\hat O_n=\hat I$ otherwise yields the joint probability $P(b_1\cdots b_k)$ of the corresponding projective subspace, so that a full bitstring can be sampled qubit by qubit, via $P(b_k|b_1\cdots b_{k-1})=P(b_1\cdots b_k)/P(b_1\cdots b_{k-1})$, with $N$ readouts instead of $2^N$. The cost is not removed: the DC term must be resolved against the combination frequencies of $2\delta$ base frequencies (Sec.~\ref{subsec:sim_circuits}), and the prefix probabilities $P(b_1\cdots b_k)$ of a spread-out distribution are themselves exponentially small. This is expected: estimating these probabilities efficiently to polynomial accuracy would allow classical hardware to reproduce the output of any quantum computation, i.e., it would imply BPP\,=\,BQP~\cite{nielsen_quantum_2010}.

\paragraph{Averaging time, precision, and the absence of quantum advantage.}  The $\mathcal{O}(\delta)$ BF count quantifies signal-generation cost only.  A complete resource accounting must include (i) the averaging time $T$ required for a target error $\gamma$, which through the small denominators $|\sum_\ell k_\ell\Omega_\ell|$ in Eq.~\eqref{eq:S_sum} grows with the density of near-resonant combination frequencies — visible already as the $\delta$-dependent prefactor in our $\delta\le4$ benchmarks, and expected to steepen for circuits with more two-qubit gates as the smallest combination frequency shrinks (Sec.~\ref{subsec:sim_circuits}); (ii) the amplitude precision and dynamic range demanded of the analog signal chain, since the information content of a $2^N$-dimensional state distributed over $2N$ signals necessarily concentrates in fine spectral and amplitude structure; and (iii) the $2^N$ readout operations above.  Taken together, these costs grow superpolynomially for general circuits — consistent with the expectation that no classical device efficiently simulates universal quantum computation.  We therefore claim \emph{no} quantum or quantum-inspired computational advantage.  The value of the architecture is that it makes the cost structure explicit and pushes all exponential overhead out of the execution stage, where analog parallelism and narrowband channels are cheap, into well-identified readout and integration-time budgets.\\

\paragraph{Scaling to large circuits.} The theoretical $\mathcal{O}(\delta)$ BF count keeps the execution stage lean, within the overall budget above. However, each BF requires a separate signal generator channel, and the analog multiplier count grows with the number of two-qubit gates.  Digital signal generation (FPGA or DDS arrays) and programmable analog front-ends are natural extensions.  A 3-qubit hardware prototype incorporating multiple layers is a near-term experimental milestone.\\

\subsection{Relation to quantum entanglement}

A natural question is how this wavebit framework relates to established quantum entanglement resource theories~\cite{horodecki_quantum_2009,chitambar_quantum_2019}.  The answer deserves stating without ambiguity.  The average in Eq.~\eqref{eq:constraint} is taken over the state \emph{vector}, i.e., over the amplitude signals themselves: a coherent demodulation of classical signals, not a physical operation on quantum states.  The physical system is classical, and its instantaneous configuration is separable at all times; a physical time-average (of density matrices) would, by convexity, remain separable.  No entanglement, in the resource-theoretic sense, is ever generated, and the framework makes no such claim.  What it provides is a \emph{dynamical parameterization} of the set of $N$-qubit amplitude vectors: a constructive, invertible map between classical multi-frequency wave dynamics and the amplitudes of arbitrary (including nonseparable) quantum states.  The reconstructed amplitude vector is, of course, a valid quantum state description (it can be tested against the Peres-Horodecki criterion~\cite{peres_separability_1996,horodecki_separability_1996} or assigned a Schmidt rank), but its entanglement is a property of the \emph{encoded data}, on the same footing as the entanglement of a state-vector stored in a digital simulator's memory.

\begin{table*}[t]
\caption{\textbf{Resources for emulating an $N$-qubit circuit with $\delta$ two-qubit gates.} Wavebit encoding, the single-signal analog emulator of La~Cour and Ott, tensor-network simulation (tw: treewidth of the circuit), and digital state-vector simulation.}
\label{tab:comparison}
\begin{ruledtabular}
\begin{tabular}{lcccc}
\rule[-11pt]{0pt}{28pt} & Channels / memory & Bandwidth & \begin{tabular}{@{}c@{}}Cost per\\ chosen amplitude\end{tabular} & Averaging time\\
\hline
\rule[-11pt]{0pt}{28pt}Wavebits &  $2N$ signals &  \begin{tabular}{@{}c@{}}$\mathcal{O}(\delta\sqrt{p_\delta})\,\omega_\mathrm{ref}$,\\ independent of $N$\end{tabular} & one $N$-way product & \begin{tabular}{@{}c@{}}$\sim1/\omega^*_\delta$,\\ up to exponential in $\delta$\end{tabular}\\
\hline
\rule[-11pt]{0pt}{28pt}Single signal~\cite{lacour_signal-based_2015} & 1 signal & $\propto 2^N$ & one tone demodulated & set by tone spacing $\propto 2^{-N}$\\
\hline
\rule[-11pt]{0pt}{28pt}Tensor network~\cite{markov_simulating_2008} & $2^{\mathcal{O}(\mathrm{tw})}$ & -- & $2^{\mathcal{O}(\mathrm{tw})}$ & --\\
\hline
\rule[-11pt]{0pt}{28pt}State vector~\cite{schrodinger_undulatory_1926} & $2^N$ amplitudes & -- & \begin{tabular}{@{}c@{}}full simulation,\\ $\mathcal{O}(2^N)$ per gate\end{tabular} & --\\
\end{tabular}
\end{ruledtabular}
\end{table*}

\subsection{Relation to prior signal-based and wave-based emulators}
\label{subsec:prior_art}

As motivated in the introduction, the idea of emulating the Hilbert-space structure of a quantum computer with classical degrees of freedom has a considerable lineage. The comparison with La~Cour and Ott's signal-based universal emulator~\cite{lacour_signal-based_2015}, however, is the most instructive (Table~\ref{tab:comparison}), as quantum amplitudes are also encoded by classical oscillatory signals. For reference, Table~\ref{tab:comparison} also lists the two digital baselines: state-vector simulation, exponential in $N$, and tensor-network contraction, exponential only in the treewidth.

In their framework, an $N$-qubit state is encoded by a single composite signal, whose $2^N$ tonals span an exponentially growing bandwidth. In contrast, our wavebit state is distributed over $2N$ signals that remain \emph{separable at every instant}: entangling correlations are never carried by any single channel but reside in the joint spectral structure of the ensemble, materializing only in the time-averaged $N$-way products of the readout stage. This shifts the exponential cost out of the execution stage entirely, since each channel remains narrowband with $\mathcal{O}(\delta)$ base frequencies overall, and concentrates it at readout and in the averaging time.

The encoding scheme also shapes the gate hardware.  In La~Cour and Ott's emulator, operating on qubit $n$ requires projection and partial-projection operators, physically realized as banks of bandpass filters that must each select $\mathcal{O}(2^N)$ distinct frequency components, and a general two-qubit operation between arbitrary qubits requires $N(N-1)$ distinct gate circuits — gate hardware that grows quadratically with register size.  In the wavebit architecture, the two basis states of each qubit occupy distinct signal channels, so qubit addressing reduces to signal routing, and a two-qubit gate between any pair applies the same two local matrices $\hat{S}_1(t)$, $\hat{S}_2(t)$ to the selected channel pairs, the pairing being carried by the shared base frequency rather than by dedicated hardware.  Gate hardware is therefore pair-agnostic and constant per entangling layer, and all-to-all connectivity costs only routing switches.

The trade is consequential in engineering terms. Bandwidth is a hard ceiling of any analog platform---Ref.~\cite{lacour_signal-based_2015} itself identifies its exponential growth as the intrinsic limit to the scalability of the single-signal approach---whereas readout operations and integration time are extensible resources: the $2^N$ readout products can be evaluated sequentially on fixed hardware, in parallel on replicated hardware, or truncated early for algorithms whose output concentrates amplitude on few basis states, as in the Grover or Deutsch--Jozsa circuits, for example. In the worst case both costs remain exponential; the wavebit architecture exchanges the less negotiable resource for the more negotiable one. The contrast is quantitative: the 40-qubit bandwidth ceiling quoted in the introduction ($0.1\,\mathrm{Hz}$--$100\,\mathrm{GHz}$) constrains a wavebit system over the same band only in the two-qubit gate count $\delta$; qubit number costs channels rather than spectrum, its ceiling being the readout budget. Moreover, the signal-based encoding of La~Cour and Ott necessitates octave frequency spacing, $\Omega_\ell\propto2^\ell$, resulting in an exponentially growing spectral range. By contrast, the present assignment $\Omega_\ell=\sqrt{p_\ell}\,\omega_{\mathrm{ref}}$ yields a maximum base frequency that scales sublinearly with the two-qubit gate count as $\Omega_{\delta}\sim\omega_{\mathrm{ref}}\sqrt{\delta\log\delta}$ by the prime number theorem~\cite{apostol_introduction_1976}, while remaining independent of the number of qubits.

Taken together, the wavebit framework provides the signal-based emulation program with an alternative encoding that eliminates the exponential bandwidth, projection, and pairwise-hardware bottlenecks of the single-signal approach, relocating the exponential cost to readout and integration time.

\subsection{Outlook}
The wavebit framework offers a constructive, gate-level route to classical analog emulation of quantum circuits, with its exponential costs explicitly localized at readout and in integration time, while providing a dynamical parameterization of nonseparable state structure rooted in multi-frequency phase coherence.

The framework places analog emulation between the digital and quantum paradigms. At fixed register size, analog signal-based emulators have been shown to outperform digital hardware: a CMOS implementation ran a six-qubit Grover search about 400 times faster than a desktop CPU and drew six times less power than an FPGA emulator at ten qubits~\cite{mourya_emulation_2023}. The wavebit encoding removes the exponential bandwidth that limits them (Sec.~\ref{subsec:prior_art} and Table~\ref{tab:comparison}). As an added benefit, any amplitude, including its phase, can be read out on demand rather than sampled, as needed to spot-check the output of a quantum processor~\cite{arute_quantum_2019}. Because the averaging time grows with the number of two-qubit gates (Sec.~\ref{subsec:sim_circuits}), this is realistic only for circuits with a moderate gate count, far below the roughly 430 two-qubit gates of Ref.~\cite{arute_quantum_2019}. The framework also suits teaching: it gives nonseparability a concrete physical form, with the tensor product enacted by multiplying signals and the nonseparable correlations carried by shared frequencies that can be observed in real time, rather than introduced as an abstract property of Hilbert-space vectors.

Regarding experimental realization, any fully connected system with programmable couplings and real-time signal routing can serve as a suitable physical substrate for wavebits, as illustrated in Fig.~\ref{fig:readout}(b). Active acoustic metamaterials, for example, naturally satisfy these requirements~\cite{padlewski_active_2023,padlewski_observation_2025,padlewski_hybrid_2024-1}.

Importantly, although the wavebit framework has been developed to emulate quantum circuits, it is not ultimately constrained by the postulates of quantum mechanics. The dynamic coefficients $\alpha_{b_n}^{(n)}(t)$ of a wavebit can be measured, copied, amplified, or fed back into the computation without disturbing the subsequent evolution, enabling mid-circuit readout, analog feedback, and signal-path branching with no direct quantum counterpart. Nonlinear operations, forbidden in quantum mechanics, are likewise implementable, in the simplest case by reading out the state, applying the nonlinear map, and re-encoding it, at the cost of the $2^N$ readout. Abrams and Lloyd showed that such operations would permit polynomial-time solutions to NP-complete and \#P problems~\cite{abrams_nonlinear_1998}; in a classical emulator, any such gain is bounded by the readout and precision costs discussed above~\cite{aaronson_np-complete_2005}, and we claim no computational advantage from them. Determining how these additional degrees of freedom can be exploited remains an exciting direction for future work.

\begin{acknowledgments}
The authors thank A.~Bossart, P.~Deymier, J.~Wang, and A.~Letcher for fruitful
discussions.\\
\end{acknowledgments}

\paragraph*{Author contributions.}
 M.M.C.V.\ contributed to the theoretical development of time-varying gate unraveling, designed and built the analog hardware prototype, conducted the experiments and manuscript preparation.  T.T.\ contributed to the theoretical analysis.  B.A.\ contributed to the theoretical framework.  H.L.\, R.F.\ supervised the project. M.P.\ conceived the nonseparability-channel formalism, led the theoretical development, simulations, manuscript preparation and supervised the project.\\

\paragraph*{Data availability.} There are no publicly available research data or software supporting this manuscript. Requests for further information or data should be sent to the authors.

\appendix

\section{Full 2-Qubit Analog Gate Coefficient Expansion}
\label{app:gate_expand}
For a general $4\times4$ gate with elements $u_{\mu\nu}$, $\mu,\nu\in\{1,2,3,4\}$,
\begin{equation}
\hat{U} = \begin{pmatrix}
u_{11} & u_{12} & u_{13} & u_{14}\\
u_{21} & u_{22} & u_{23} & u_{24}\\
u_{31} & u_{32} & u_{33} & u_{34}\\
u_{41} & u_{42} & u_{43} & u_{44}
\end{pmatrix},
\end{equation}
the two time-dependent $2\times2$ matrices $\hat{S}_1(t)$ and $\hat{S}_2(t)$ whose time-averaged tensor product yields $\hat{U}$ are given as follows.

Choose 16 orthonormal signals $\{s_k(t)\}_{k=1}^{16}$ satisfying $\avginf{s_k s_{k'}^*}=\delta_{kk'}$, with $\delta_{kk'}$ the Kronecker delta, which can be understood as a usual inner product. A natural choice is $s_k(t) = e^{\mathrm{i}k\Omega t}$, $k=1,\ldots,16$. For readability, we drop the time dependence of these signals in this appendix, $s_k\equiv s_k(t)$.

Define:
\begin{equation}
\hat{S}_1(t) = \begin{pmatrix}a_1(t) & b_1(t)\\c_1(t) & d_1(t)\end{pmatrix},\quad
\hat{S}_2(t) = \begin{pmatrix}a_2(t) & b_2(t)\\c_2(t) & d_2(t)\end{pmatrix},
\end{equation}
where
\begin{align}
a_1(t) &= \sqrt{u_{11}}s_1 + \sqrt{u_{12}}s_2 + \sqrt{u_{21}}s_3 + \sqrt{u_{22}}s_4,\\
b_1(t) &= \sqrt{u_{13}}s_5 + \sqrt{u_{14}}s_6 + \sqrt{u_{23}}s_7 + \sqrt{u_{24}}s_8,\\
c_1(t) &= \sqrt{u_{31}}s_9 + \sqrt{u_{32}}s_{10} + \sqrt{u_{41}}s_{11} + \sqrt{u_{42}}s_{12},\\
d_1(t) &= \sqrt{u_{33}}s_{13} + \sqrt{u_{34}}s_{14} + \sqrt{u_{43}}s_{15} + \sqrt{u_{44}}s_{16},
\end{align}
and
\begin{align}
a_2(t) &= \sqrt{u_{11}}s_1^* + \sqrt{u_{13}}s_5^* + \sqrt{u_{31}}s_9^* + \sqrt{u_{33}}s_{13}^*,\\
b_2(t) &= \sqrt{u_{12}}s_2^* + \sqrt{u_{14}}s_6^* + \sqrt{u_{32}}s_{10}^* + \sqrt{u_{34}}s_{14}^*,\\
c_2(t) &= \sqrt{u_{21}}s_3^* + \sqrt{u_{23}}s_7^* + \sqrt{u_{41}}s_{11}^* + \sqrt{u_{43}}s_{15}^*,\\
d_2(t) &= \sqrt{u_{22}}s_4^* + \sqrt{u_{24}}s_8^* + \sqrt{u_{42}}s_{12}^* + \sqrt{u_{44}}s_{16}^*.
\end{align}

\textit{Verification.}  Consider the $(1,1)$ element of $\avginf{\hat{S}_1(t)\otimes\hat{S}_2(t)}$,
which is $\avginf{a_1(t)\cdot a_2(t)}$.  By orthonormality $\avginf{s_k s_{k'}^*}=\delta_{kk'}$, all cross-products vanish and only the $s_1 s_1^*$ term survives:
\begin{equation}
    {\avginf{a_1(t)a_2(t)} = \sqrt{u_{11}}\sqrt{u_{11}}\avginf{s_1 s_1^*} = u_{11}.}
\end{equation}
Each element of $\hat{U}$ is recovered analogously; the complete tensor product expansion over all 16 matrix elements follows by the same argument.

\section{Frequency Independence of Time Averages and Irrational Frequencies}
\label{app:thm1}
\begin{theorem}[Frequency independence of time averages]
\label{thm:freq_indep}
Let $A_1(t),\ldots,A_\delta(t)\in\mathbb{C}^{m\times m}$ be continuously
differentiable, periodic matrices with fundamental frequencies
$\Omega_1,\ldots,\Omega_\delta$ that are linearly independent over $\mathbb{Q}$.
Then
\begin{equation}
     \avginf{A_1(t)}\cdots\avginf{A_\delta(t)} = \avginf{A_1(t)\cdots A_\delta(t)}.
    \label{eq:freq_indep}
\end{equation}
\end{theorem}
\begin{proof}
Expand each matrix in its Fourier series:
\begin{equation}
    A_\ell(t) = \sum_{k=-\infty}^{\infty} \hat{A}_{\ell,k}\,e^{\mathrm{i}k\Omega_\ell t}.
    \label{eq:C_fourier}
\end{equation}
The product becomes
\begin{align}
    &A_1(t)\cdots A_\delta(t) \notag\\
    &= \sum_{k_1,\ldots,k_\delta\in\mathbb{Z}}
    \hat{A}_{1,k_1}\cdots\hat{A}_{\delta,k_\delta}\,
    e^{\mathrm{i}(k_1\Omega_1+\cdots+k_\delta\Omega_\delta)t}.
    \label{eq:C_product}
\end{align}
The time average $\avginf{e^{\mathrm{i}\omega t}}$ equals 1 if $\omega=0$ and 0 otherwise.  By linear independence over $\mathbb{Q}$, the sum $k_1\Omega_1+\cdots+k_\delta\Omega_\delta=0$ if and only if $k_1=\cdots=k_\delta=0$. Therefore
\begin{equation}
\begin{aligned}
    \avginf{A_1(t)\cdots A_\delta(t)} &= \hat{A}_{1,0}\cdots\hat{A}_{\delta,0}\\
    &= \avginf{A_1(t)}\cdots\avginf{A_\delta(t)}.
\end{aligned}
\end{equation}
\end{proof}

\begin{lemma}[Rational frequencies are insufficient]
\label{lem:rational}
No set of $n\geq 2$ rational numbers $\{\Omega_\ell\}\subset\mathbb{Q}$ is
linearly independent over $\mathbb{Q}$.  Therefore, BFs must be chosen from
irrational (or transcendental) frequencies.
\end{lemma}

\begin{proof}
Let $\Omega_1=p_1/q_1$ and $\Omega_2=p_2/q_2$ be two elements of the set, with $p_\ell,q_\ell\in\mathbb{Z}$ and $q_\ell>0$. Then $(p_2q_1)\,\Omega_1-(p_1q_2)\,\Omega_2=0$, a non-trivial integer linear combination unless $p_1=p_2=0$, in which case $1\cdot\Omega_1=0$ is. Hence the set is linearly dependent over $\mathbb{Q}$.
\end{proof}

\noindent\textbf{Practical frequency assignment.}  Irrational BFs can be obtained using prime numbers --- specifically the square roots of distinct primes times a reference frequency : $\Omega_\ell = \sqrt{p_\ell}\,\omega_\mathrm{ref}, \quad p_\ell\text{ the }\ell\text{-th prime}$. Because the square root of a prime is irrational, $\{\Omega_1,\Omega_2,\ldots\}=\{\sqrt{2},\sqrt{3},\sqrt{5},\sqrt{7},\ldots\}\,\omega_\mathrm{ref}$ are linearly independent over $\mathbb{Q}$ by Besicovitch's theorem~\cite{besicovitch_linear_1940}.  Another advantage of this assignment is given by the prime number theorem, which states that primes scale as \(p_\ell \sim \ell\log\ell\) for large \(\ell\)~\cite{apostol_introduction_1976}. Consequently, in a large hardware implementation, the highest layer frequency grows only as $\sqrt{p_\delta} \sim \sqrt{\delta\ln\delta}$ — polynomial rather than exponential — keeping the digital sampling cost manageable. A comparison with alternative frequency schemes is given in the Supplemental Material~\cite{ref_2_supp}, Sec.~S5.

\section{Layer decomposition and circuit universality}
\label{app:circuit_univ}
We wish to emulate an arbitrary $N$-qubit unitary quantum circuit by unraveling all of its gates. Any such circuit can be decomposed into 2-qubit gates from a universal gate set $\Lambda$ (e.g., Clifford + T-gate)~\cite{lloyd_almost_1995,boykin_universal_1999}. So we assume the circuit we wish to emulate has been constructed exclusively using single-qubit and 2-qubit gates.

We define an \emph{entangling layer}, or simply \emph{layer}, as a single two-qubit unitary gate embedded in the $N$-qubit Hilbert space,
\begin{equation}
    \hat{G}_\ell
    =
    \hat{U}_{\ell}^{(n,n')},
    \label{eq:layer}
\end{equation}
where
\begin{equation}
    \hat{U}_{\ell}^{(n,n')}
    =
    P_{n,n'}^{-1}
    \left(
    \hat{U}_{\ell}
    \otimes
    \hat{I}^{\otimes(N-2)}
    \right)
    P_{n,n'},
    \label{eq:U_nn}
\end{equation}
with $P_{n,n'}$ the permutation operator that reorders the tensor factors so that qubits $n$ and $n'$ occupy the first two tensor factors and $\hat{U}_{\ell}$ is the $4\times4$ unitary operator unraveled in Sec.~\ref{subsec:2qubit_gate}. Consequently, $\hat{U}_{\ell}^{(n,n')}$ acts as $\hat{U}_{\ell}$, on qubits $n$ and $n'$, and as the identity on all remaining qubits. Single-qubit gates are not layers: they are applied between layers as static matrices (see below).
  \\

In each layer, $\hat{U}_{\ell}$ is a genuinely entangling
    2-qubit gate, unraveled into time-varying wavebit signals at its own
    \emph{base frequency} $\Omega_\ell$. Single-qubit gates, in contrast, are applied as static $2\times2$ matrices to the channels of their wavebit (Sec.~\ref{subsec:1qubit_gate}); they require no dynamic signals and no base frequency, and are not passed through the construction of Eq.~\eqref{eq:S1_elements}.
The \emph{two-qubit gate count} $\delta$ of a circuit counts its entangling layers, equal to the number of distinct base frequencies required. A circuit with $\delta$ layers is $\hat{G}=\hat{V}_\delta\hat{G}_\delta\hat{V}_{\delta-1}\cdots\hat{V}_1\hat{G}_1\hat{V}_0$, where each $\hat{V}_\ell$ is a (possibly trivial) product of single-qubit gates.\\

Each layer $\hat{G}_\ell$ is unraveled as
\begin{equation}
\begin{aligned}
    \hat{G}_\ell
    &= \hat{U}_{\ell}^{(n,n')} \\
    &= P_{n,n'}^{-1}
    \left(
        \hat{U}_{\ell}
        \otimes
        \hat{I}^{\otimes(N-2)}
    \right)
    P_{n,n'} \\
    &= P_{n,n'}^{-1}
    \left(
        \avginf{\hat{U}_{\ell}(t)}
        \otimes
        \hat{I}^{\otimes(N-2)}
    \right)
    P_{n,n'} \\
    &\coloneqq \avginf{\hat{G}_\ell(t)},
\end{aligned}
\label{eq:layer_unravel}
\end{equation}

where $ \hat{U}_{\ell}(t) = (\hat{S}_1(t)\otimes\hat{S}_2(t))_\ell$ (Eq.~\eqref{eq:gate_unravel}). This layer uses orthonormal signals built from harmonics of a layer-specific base frequency $\Omega_\ell$: $s_{\ell,k}(t)=e^{\mathrm{i}k\Omega_\ell t}$, $k=1,\ldots,16$.  Single-qubit gates are constant matrices that factor out trivially and impose no frequency constraint. The $\mathbb{Q}$-independence requirement therefore applies only to the set of base frequencies of the layers: no base frequency may be expressed as a rational linear combination of the others.  Under this condition, Theorem~\ref{thm:freq_indep} guarantees that the time averages factorize across all layers, giving the main circuit emulation result:
\begin{equation}
    \hat{G}\ket{\Psi}
    = \avginf{\hat{V}_\delta\hat{G}_\delta(t)\hat{V}_{\delta-1}\cdots\hat{V}_1\hat{G}_1(t)\hat{V}_0}\ket{\Psi},
    \label{eq:circuit_unravel}
\end{equation}
where $\ket{\Psi}$ is any separable input state.  The entire multi-qubit circuit is reproduced as a single time average of a product of separable time-dependent operators.\\

\begin{figure}[htb]
\centering
\includegraphics{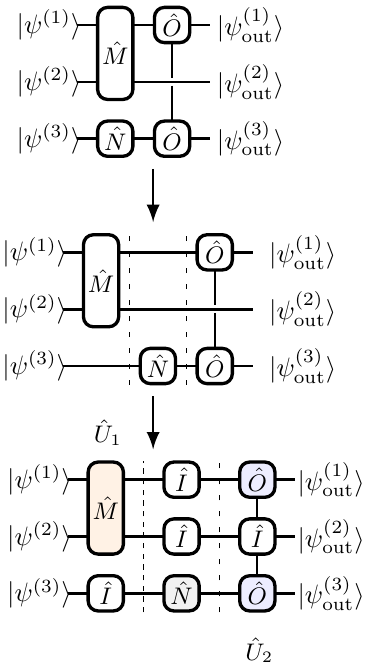}
\caption{\textbf{Layer decomposition for a 3-qubit circuit.} Starting from the sequential gate application (\emph{top}), each two-qubit gate defines one entangling layer (\emph{middle}), and the decomposition is made explicit by color (\emph{bottom}). Layers $\ell=1$ ($\hat{M}$, orange, acting on $|{\psi^{(1)}}\rangle$ and $|{\psi^{(2)}}\rangle$) and $\ell=2$ ($\hat{O}$, blue, acting on $|{\psi^{(1)}}\rangle$ and $|{\psi^{(3)}}\rangle$) contain genuinely entangling 2-qubit gates and each require a distinct base frequency $\Omega_\ell$. Since $\hat{O}$ couples non-adjacent qubits, it is drawn as two linked single-qubit blocks rather than a single contiguous box. The single-qubit gate $\hat{N}$ (gray) between them is not an entangling layer: it is applied as a static matrix to $|{\psi^{(3)}}\rangle$, with no base frequency. The two-qubit gate count is $\delta=2$. By Theorem~\ref{thm:freq_indep}, the time averages of the two layers factorize, reproducing the full circuit as a product of independent local time averages.}
\label{fig:sublayer_decomp}
\end{figure}

For concreteness, Fig.~\ref{fig:sublayer_decomp} illustrates this for a 3-qubit example based on the sandwich circuit presented in the main text (Sec.~\ref{subsec:sandwich}). Layers $\hat{G}_1$ ($\hat{M}$) and $\hat{G}_2$ ($\hat{O}$) contain genuinely entangling 2-qubit gates and each require
a distinct $\mathbb{Q}$-independent base frequency.  Consequently, the two-qubit gate count is $\delta=2$.\\

\noindent\textit{Required number of base frequencies.} Each layer requires a layer-specific base frequency (BF); its harmonics build the signals of $\hat{S}_1(t)$ and $\hat{S}_2(t)$. Single-qubit gates require none. The total number of BFs, $N_\mathrm{BF}$, is thus equal to the two-qubit gate count $\delta$. Crucially, $N_\mathrm{BF}$ is far smaller than the number of NSCs: during the \emph{readout stage}, the $N$ wavebit signals are multiplied together, and this $N$-way product automatically generates all $2^N$ distinct combination frequencies — each carrying one target state coefficient $\alpha_\mathbf{b}$.  The NSCs therefore emerge from the BF signals at readout, not from additional signal sources; their exponential count reflects the dimensionality of the output state, not a cost of the circuit execution.\\

\noindent\emph{Summary.} The BF count then depends only on the two-qubit gate count: $N_\mathrm{BF} = \mathcal{O}(\delta)$, independent of system size $N$. Consequently, $N_\mathrm{BF}\ll 2^N$ — the BF count is polynomial while the NSC count grows exponentially — reflecting the separation between circuit execution cost and output-state dimensionality.\\

\section{Proof of the Finite-Time Error Bound}
\label{app:thm2}
Considering a set of irrational (or transcendental) frequencies, let $S = \sum_{\mathbf{k}\neq\mathbf{0}}
\|\hat{A}_{1,k_1}\cdots\hat{A}_{\delta,k_\delta}\|/|\sum_\ell k_\ell\Omega_\ell|$.
If $S$ converges, then
$\|\avginf{A_1\cdots A_\delta}-\avgT{A_1\cdots A_\delta}\|\leq 2S/T$.

\begin{proof}
Define $P(t) = A_1(t)\cdots A_\delta(t)$.  From Eq.~\eqref{eq:C_product},
\begin{equation}
    P(t) = \hat{P}_0 + \sum_{\mathbf{k}\neq\mathbf{0}}
    \hat{P}_\mathbf{k}\,e^{\mathrm{i}\omega_\mathbf{k} t},
\end{equation}
where $\hat{P}_\mathbf{k}=\hat{A}_{1,k_1}\cdots\hat{A}_{\delta,k_\delta}$ and $\omega_\mathbf{k}=\sum_\ell k_\ell\Omega_\ell\neq0$ for $\mathbf{k}\neq\mathbf{0}$. The finite-time average gives
\begin{equation}
    \avgT{P(t)} = \hat{P}_0 + \sum_{\mathbf{k}\neq\mathbf{0}}
    \hat{P}_\mathbf{k}\,\frac{e^{\mathrm{i}\omega_\mathbf{k} T}-1}{\mathrm{i}\omega_\mathbf{k} T}.
\end{equation}
Using $|e^{\mathrm{i}\omega T}-1|\leq 2$ and the sub-multiplicative norm:
\begin{align}
    \|\avginf{P}-\avgT{P}\| &= \left\|\sum_{\mathbf{k}\neq\mathbf{0}}
    \hat{P}_\mathbf{k}\,\frac{e^{\mathrm{i}\omega_\mathbf{k} T}-1}{\mathrm{i}\omega_\mathbf{k} T}\right\|\\
    &\leq \sum_{\mathbf{k}\neq\mathbf{0}}\|\hat{P}_\mathbf{k}\|\,
    \frac{2}{|\omega_\mathbf{k}|T}
    = \frac{2S}{T}.
\end{align}
\end{proof}

The bound can be minimized by choosing bandlimited signals (so that high-$k$ Fourier components are zero and $S$ converges rapidly) and maximizing the frequency separations $|\omega_\mathbf{k}|$.\\

\section{Direct N-Qubit Unraveling Formula}
\label{app:gate_unravel}
Here, we wish to formalize a systematic way for unraveling an arbitrary $N$-qubit state. Following Eq.~\eqref{eq:N_qubit_dyn}, we parametrize each dynamic amplitude using $2^N$ NSCs:
\begin{equation}
    \alpha_{b_n}^{(n)}(t) = \sum_{\mathbf{u}\in\{0,1\}^N}
    \beta^{(n)}_{b_n}[\mathbf{u}]\, e^{\mathrm{i}\Omega^{(n)}_{b_n}[\mathbf{u}]\,t}.
    \label{eq:alpha_expansion}
\end{equation}
Inserting Eq.~\eqref{eq:alpha_expansion} into the constraint~\eqref{eq:constraint}, the bracket $\avginf{e^{\mathrm{i}\sum_n\Omega^{(n)}_{b_n}[\mathbf{u}^{(n)}]t}}$ vanishes unless $\sum_n\Omega^{(n)}_{b_n}[\mathbf{u}^{(n)}]=0$.  Frequencies are chosen so this sum vanishes \emph{if and only if} $\mathbf{u}^{(1)}=\cdots=\mathbf{u}^{(N)}=\mathbf{b}$. This is achieved by (i) assigning random irrational-ratio frequencies to qubits $1,\ldots,N-1$, and (ii) retuning the last qubit:
\begin{equation}
    \Omega^{(N)}_{u_N}[\mathbf{u}] = -\sum_{n=1}^{N-1}\Omega^{(n)}_{u_n}[\mathbf{u}],
    \quad\forall\,\mathbf{u}\in\{0,1\}^N.
    \label{eq:retune}
\end{equation}
Setting $\beta^{(n)}_{b_n}[\mathbf{b}] = \alpha_\mathbf{b}^{1/N}$ then yields
\begin{equation}
    \alpha_\mathbf{b} = \prod_{n=1}^N \beta^{(n)}_{b_n}[\mathbf{b}],
    \label{eq:coef_result}
\end{equation}
completing the construction. The obvious limitation of this direct approach is that it requires a large number of frequencies and amplitudes to be specified and thus stored in memory, making it computationally intensive for large $N$.

The direct $N$-qubit simulation in Sec.~\ref{subsec:sim_states} uses base frequencies $\Omega_n = \sqrt{p_n}\,\omega_\mathrm{ref}$ for the first $N-1$ qubits and last-qubit retuning via Eq.~\eqref{eq:retune}.

\bibliography{refs_merged,ref_2_supp}

\pdfximage{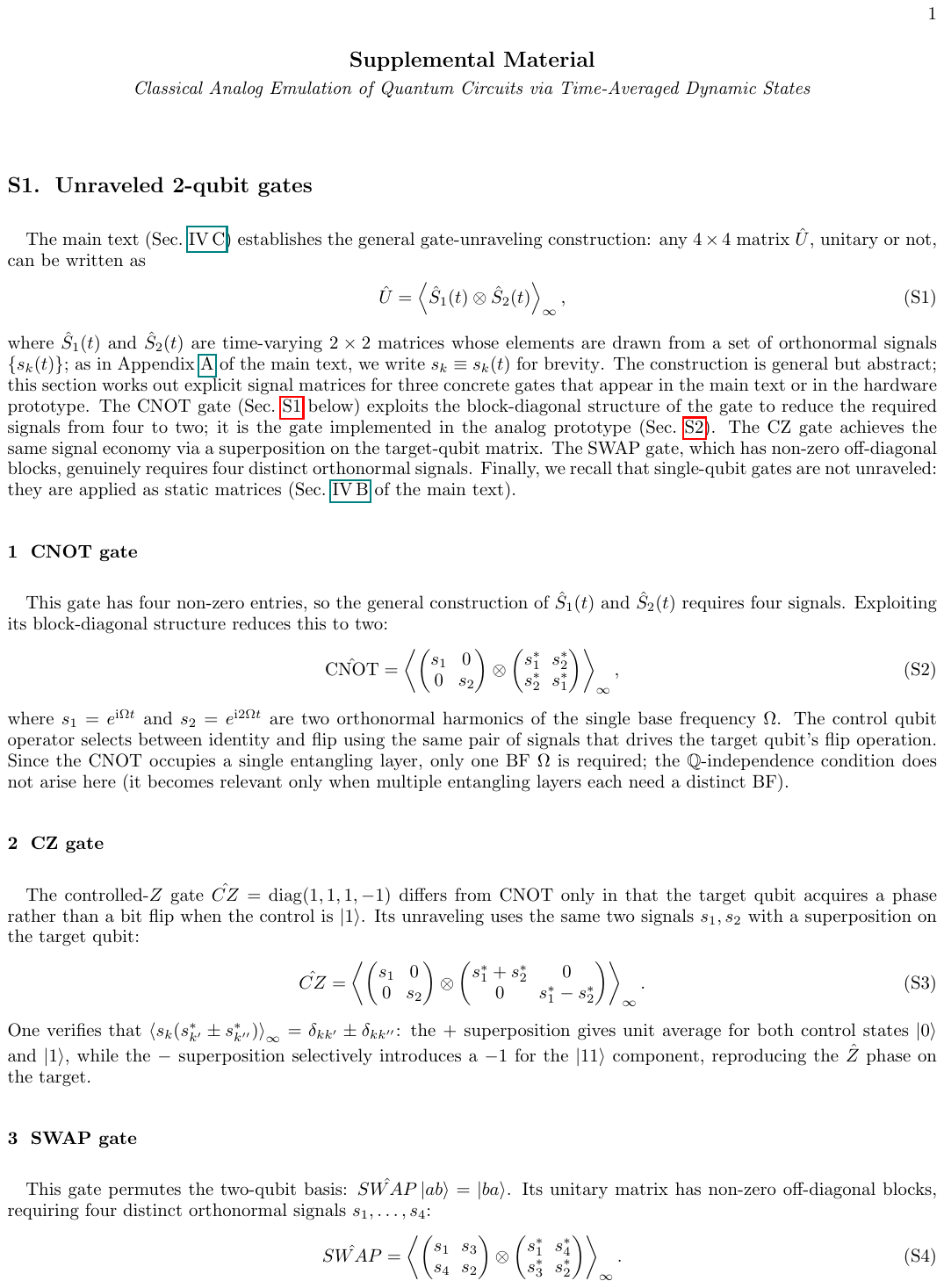}
\foreach \x in {1,...,\the\pdflastximagepages}{%
    \clearpage
    \includepdf[pages={\x},pagecommand={\thispagestyle{empty}}]{prapplied_supp.pdf}
}

\end{document}